\documentclass[journal]{IEEEtran}
\usepackage{amsmath}
\usepackage{multirow}
\usepackage{graphicx}
\usepackage{cite}
\usepackage{subfigure}
\usepackage{amssymb}
\usepackage{algorithm}
\usepackage{algorithmic}

\ifCLASSINFOpdf
\else
\fi
\begin{document}

\title{Improving the Adversarial Robustness for Speaker Verification by Self-Supervised Learning}

\author{Haibin Wu*,
        Xu Li*,
        Andy T. Liu,
        Zhiyong Wu,
        Helen Meng~\IEEEmembership{Fellow,~IEEE,}
        Hung-yi Lee
\thanks{Manuscript received April 19, 2021; revised August 12, 2021 and October 10, 2021; accepted November 14, 2021. Date of publication December 10, 2021; date of current version January 8, 2022. We thank the National Center for High-performance Computing (NCHC) of National Applied Research Laboratories (NARLabs) in Taiwan for providing computational and storage resources. This work was partially supported by the Ministry of Science and Technology of Taiwan under Project No. 110-2628-E-002-001, HKSAR Government’s Research Grants Council General Research Fund (ref. no. 14208718) and the Centre for Perceptual and Interactive Intelligence, a CUHK InnoCentre. The associate editor coordinating the review of this manuscript and approving it for publication was Dr. Jing Huang. (Haibin Wu and Xu Li contributed equally to this work.) (Corresponding author: Hung-Yi Lee.)}
\thanks{Haibin Wu, Andy T. Liu and Hung-yi Lee are with the Graduate Institute of Communication Engineering, National Taiwan University, Taiwan (e-mail: \{f07921092, f07942089, hungyilee\}@ntu.edu.tw).}
\thanks{Xu Li and Helen Meng are with the Department of Systems Engineering and Engineering Management, The Chinese University of Hong Kong, Hong Kong (email: \{xuli,hmmeng\}@se.cuhk.edu.hk).}
\thanks{Zhiyong Wu is with the Shenzhen International Graduate School, Tsinghua University (email: zywu@se.cuhk.edu.hk).}
}

\markboth{Journal of \LaTeX\ Class Files,~Vol.~14, No.~8, August~2015}%
{Shell \MakeLowercase{\textit{et al.}}: Bare Demo of IEEEtran.cls for IEEE Journals}

\maketitle

\begin{abstract}
Previous works have shown that automatic speaker verification (ASV) is seriously vulnerable to malicious spoofing attacks, such as replay, synthetic speech, and recently emerged adversarial attacks.
Great efforts have been dedicated to defending ASV against replay and synthetic speech; however, only a few approaches have been explored to deal with adversarial attacks.
All the existing approaches to tackle adversarial attacks for ASV require the knowledge for adversarial samples generation, but it is impractical for defenders to know the exact attack algorithms that are applied by the in-the-wild attackers.
This work is among the first to perform adversarial defense for ASV without knowing the specific attack algorithms.
Inspired by self-supervised learning models (SSLMs) that possess the merits of alleviating the superficial noise in the inputs and reconstructing clean samples from the interrupted ones, this work regards adversarial perturbations as one kind of noise and conducts adversarial defense for ASV by SSLMs.
Specifically, we propose to perform adversarial defense from two perspectives: 1) adversarial perturbation purification and 2) adversarial perturbation detection.
The purification module aims at alleviating the adversarial perturbations in the samples and pulling the contaminated adversarial inputs back towards the decision boundary.
Experimental results show that our proposed purification module effectively counters adversarial attacks and outperforms traditional filters from both alleviating the adversarial noise and maintaining the performance of genuine samples.
The detection module aims at detecting adversarial samples from genuine ones based on the statistical properties of ASV scores derived by a unique ASV integrating with different number of SSLMs.
Experimental results show that our detection module helps shield the ASV by detecting adversarial samples.
Both purification and detection methods are helpful for defending against different kinds of attack algorithms.
Moreover, since there is no common metric for evaluating the ASV performance under adversarial attacks, this work also formalizes evaluation metrics for adversarial defense considering both purification and detection based approaches into account.
We sincerely encourage future works to benchmark their approaches based on the proposed evaluation framework.

\end{abstract}


%
\IEEEpeerreviewmaketitle

\ifCLASSOPTIONcaptionsoff
  \newpage
\fi

\section{Introduction}
\label{sec:introduction}
%
%
%
%
\IEEEPARstart{A}{utomatic} speaker verification (ASV) deals with a task of certifying whether a certain piece of utterance belongs to a given speaker. It has been intensively studied for decades, and a large variety of cutting-edge ASV models have been proposed \cite{mak2020machine}. These ASV models can be classified into three most representative frameworks: 1) i-vector based speaker embedding systems \cite{dehak2010front,kenny2012small,prince2007probabilistic,garcia2011analysis,lei2014novel}, 2) deep neural network (DNN) based speaker embedding systems \cite{variani2014deep,snyder2017deep,snyder2018x,li2020bayesian}, and 3) end-to-end speaker verification systems \cite{zhang2016end,heigold2016end,snyder2016deep}. ASV is one of the most essential technology for biometric identification and is widely used in real-world applications, including smartphones, banking systems, IoT devices, etc.

Given that such applications are security-critical, the robustness of ASV systems is of high relevance. However, recent works have shown that ASV systems are vulnerable to malicious attacks, such as impersonation \cite{zetterholm2004comparison,wu2015spoofing}, replay \cite{wu2015spoofing,wu2014study}, synthetic speech \cite{shchemelinin2013examining,shchemelinin2014vulnerability,kinnunen2012vulnerability} and very lately emerged adversarial attacks \cite{das2020attacker,kreuk2018fooling}. These attacks can pose severe threats to the state-of-the-art (SOTA) ASV systems. To tackle this issue, ASVspoof challenge series \cite{yamagishi2019asvspoof, kinnunen2017asvspoof, wu2015asvspoof} have been held to develop strong countermeasures \cite{gomez2020kernel,lavrentyeva2019stc,li2020replay,li2021channel} mainly against replay and synthetic speech. Due to the very late emergence of adversarial attacks, the defense approaches against them are rare and limited, which is the focus of this paper.

The concept of adversarial attack was firstly proposed by \cite{szegedy2013intriguing}.
They slightly modified the genuine clean samples by deliberately crafted tiny perturbations to derive adversarial counterparts, which will fool image classification models with super-human performance embarrassingly predict wrong results. 
This shortcoming of image classification models is particularly surprising because the modiﬁcations are often imperceptible, or barely perceptible, to humans.

Not only computer vision models but also audio processing models are subject to adversarial attacks.
Carlini and Wagner \cite{carlini2018audio} successfully conduct adversarial attacks against the DeepSpeech \cite{hannun2014deep}, a SOTA automatic speech recognition (ASR) neural network model.
Given any piece of audio, whether speech, silence, or music, the authors can perform adversarial attacks to craft another piece of audio, which is over 99\% similar to the original one, but can be predicted by the ASR model as any transcription predefined by the authors with 100\% success rate.
\cite{yuan2018commandersong,schonherr2018adversarial,yakura2018robust,taori2019targeted,qin2019imperceptible,cisse2017houdini,iter2017generating,alzantot2018did} also do adversarial attacks towards ASR from different aspects.
Existing works also show the application of adversarial attacks for other speech processing tasks, including anti-spoofing for ASV \cite{liu2019adversarial,wu2020defense_2,zhang2020black,wu2020defense,kassis2021practical}, music classification \cite{kereliuk2015deep}, voice conversion \cite{huang2021defending}, sound event classification \cite{subramanian2019robustness} and source separation \cite{takahashi2020adversarial}.

Also, previous works \cite{kreuk2018fooling,li2020adversarial,xie2020real,marras2019adversarial,li2020practical,chen2019real,gong2017crafting,wang2020inaudible,villalba2020x,das2020attacker,kassis2021practical} showed that adversarial attacks can cause catastrophic performance drop of the SOTA ASV systems, which undermine the deployment of ASV systems in the real-world.
\cite{kreuk2018fooling} is among the first ones to investigate the vulnerability of ASV models to adversarial attacks.
They demonstrate that the end-to-end text-dependent ASV models can be fooled by malicious attackers both in cross-dataset and cross-feature scenarios.
Some early works attack ASV systems with few speakers \cite{gong2017crafting,chen2019real}.
What's more, SOTA ASV models, including i-vector and x-vector systems, are also vulnerable to adversarial attacks.
\cite{villalba2020x} benchmarks the adversarial robustness of three x-vector systems, which are the SOTA models in recent ASV evaluations, and shows they are subject to adversarial attacks.
Li et al. \cite{li2020adversarial} illustrate the i-vector system can be fooled by adversarial attacks and the adversarial samples crafted from i-vector systems attain the transferability to x-vector systems.
\cite{kassis2021practical} firstly proposes to attack both the anti-spoofing and the ASV as a whole system.
In order to attack real-world applications possibly, attackers also pour great efforts to make adversarial attacks over-the-air, inaudible and universal.
Instead of feeding the adversarial examples directly into the ASV systems, Li et al. \cite{li2020practical} first demonstrate that the x-vector system is vulnerable to over-the-air adversarial attacks in the physical world.
\cite{marras2019adversarial} explores the existence of master voices, i.e. adversarial utterances which can mimic a large number of users by an adversarial optimization method.
Wang et al. \cite{wang2020inaudible} introduce the psychoacoustic principle \cite{lin2015principles} of frequency masking to generate more inaudible adversarial perturbations.
The experimental results on AIshell-1 dataset \cite{bu2017aishell} show that their proposed adversarial attack method can manipulate the x-vector based speaker recognition system into identifying a piece of audio, whether speech or music, as any target (i.e. adversary-desired) speaker.
Xie et al. \cite{xie2020real} propose the first audio-agnostic, real-time, over-the-air adversarial perturbation against the x-vector based ASV systems.
With the pervasiveness of ASV systems in safety-critical environment, mitigating the vulnerability of ASV models to adversarial attacks is of high priority.

However, how to conduct adversarial defense effectively for ASV systems still remains an open question. 
Previous attempts \cite{wang2019adversarial,li2020investigating,zhang2020adversarial} require knowledge about attack methods' details during adversarial sample generation.
It is impractical that the defenders know exactly the attack algorithms which the in-the-wild attackers will choose in advance.

We make the first attempt to conduct adversarial defense for ASV by self-supervised learning without the knowledge of the attack algorithms. 
Self-supervised learning has aroused keen attention recently, and the generative approach based on it possesses the merits of alleviating the superficial noise in the inputs and extracting the pivotal information from the contaminated inputs after training.
To some extent, adversarial perturbation is also a kind of noise, so we propose a defense framework based on self-supervised learning based models.
In this work, the self-supervised learning based model is adopted to purify the adversarial perturbation and reform the adversarial samples, which is called self-supervised learning based reformer (SSLR).
The defense framework consists of an adversarial perturbation purification module which attempts to pull the contaminated adversarial inputs back towards the decision boundary based on the purification ability of the SSLR, and an adversarial perturbation detection module that aims at directly detecting the adversarial samples.

Early works about adversarial attack \cite{kreuk2018fooling, gong2017crafting, chen2019real, xie2020real,li2020practical} and defense \cite{wang2019adversarial,zhang2020adversarial} focus on speaker classiﬁcation tasks with few speakers.
In this work, we focus on the speaker verification task, which is an open-set problem and more challenging than close-set classification tasks.
We adopt Voxceleb1 \cite{nagrani2017voxceleb} as our benchmark dataset, which includes 148,642 short clips of human speech from 1,251 speakers.
Our experimental results on Voxceleb1 demonstrate that the proposed framework helps counter adversarial attacks for SOTA ASV models from both purification and detection perspectives.

Our contributions are as follows:
\begin{itemize}
    \item This paper is among the first ones to perform adversarial defense for ASV models without knowing the adversarial sample generation process. As the beginning work of this direction, there is no baseline work for reference. So we firstly utilize traditional filters as countermeasures for adversarial attacks and employ them as our baseline.
    \item The proposed adversarial defense framework based on self-supervised learning takes adversarial perturbation purification and detection into account. It effectively improves the robustness of SOTA ASV models against adversarial attacks. Specifically, the adversarial perturbation purification module effectively counters adversarial attacks and outperforms traditional filters from both alleviating the adversarial noise and maintaining the performance of genuine samples. Also, the adversarial perturbation detection module helps shield the ASV by detecting the adversarial samples with an accuracy of around 80\% against different attack algorithms.
    \item We also firstly formalize the evaluation metrics for adversarial defense on ASV, as shown in Section~\ref{sec:evaluation-metrics}, and sincerely encourage future works to benchmark their approaches based on the proposed evaluation metrics.
\end{itemize}

\section{Related work}
Existing approaches to protect ASV systems against adversarial attacks can be divided into two categories: purification methods and detection methods.

Purification methods regard the adversarial perturbations within modified samples as a particular kind of noise. 
They tackle this problem by either developing some filters \cite{hu2018squeeze} in front of ASV systems to purify the samples or leveraging adversarial training \cite{goodfellow2014explaining,szegedy2013intriguing} to make ASV systems themselves robust against this kind of noise. 
After applying purification methods, adversarial samples become less offensive, and thus given the adversarial samples, ASV can give the same predictions as the corresponding genuine ones.
Wang et al. \cite{wang2019adversarial} and Wu et al. \cite{wu2020defense} harness the idea of adversarial training to alleviate the vulnerability of ASV models and anti-spooﬁng models for ASV, respectively.
\cite{wang2019adversarial} mixes the adversarial objective function with the classification objective as regularization while \cite{wu2020defense} injects adversarial samples into training set for data augmentation. 
Adversarial training requires the knowledge of the attack methods for adversarial sample generation and is sensitive to specific attack algorithms.
However, it does not make sense that the system designers know the exact attack algorithms to be adopted by attackers.
Zhang et al. \cite{zhang2020adversarial} propose to train an independent DNN filter and apply it before the ASV system to mitigate the adversarial perturbations and purify the adversarial samples.
However, this method \cite{zhang2020adversarial} requires adversarial samples to train the filtering module, so the defense method is specific and even over-fitting to the attack algorithms for generating the training samples.
\cite{wu2020defense} equips the anti-spoofing models for ASV with hand-crafted filters, including Gaussian filter, median filter and mean filter, to counter the adversarial attacks.
Hand-crafted filters haven't been used for adversarial defense on ASV.
We adopt the above filters to protect ASV systems and compare their performance with our proposed method.

Detection methods aim at distinguishing adversarial samples from genuine ones.
The detected adversarial samples are discarded and ASV systems refuse to make prediction on them.
\cite{li2020investigating} proposes a detection network for ASV to distinguish the adversarial samples from genuine samples.
However, training such a detection network also needs to know the specific attack algorithms to generate adversarial samples.

To our best knowledge, only a few approaches \cite{wang2019adversarial,li2020investigating,zhang2020adversarial} were proposed to protect ASV systems against adversarial attacks, and how to effectively purify and mitigate the adversarial perturbations for ASV systems is still an open question.
In this work, we propose to defend ASV systems against adversarial attacks by the SSLR models.
Specifically, the proposed defense framework is composed of an adversarial purification module and an adversarial perturbation detection module, which mitigates the adversarial noise or directly detects the adversarial samples, respectively.
What's more, compared with previous works \cite{wang2019adversarial,li2020investigating,zhang2020adversarial}, our method doesn't require the knowledge of adversarial samples generation process in advance.
Experimental results show that our proposed method can effectively protect ASV systems against adversarial attacks from both purification and detection perspectives.

Our previous work \cite{wu2021adversarial} adopts the self-supervised learning based model as a deep filter to alleviate adversarial attack for the r-vector system. 
In contrast to \cite{wu2021adversarial}, we further investigate the potential of self-supervised learning based models against adversarial attack for ASV from different perspectives:
\begin{itemize}
\item \cite{wu2021adversarial} merely considers self-supervised learning models for purification use, yet this work harnesses it for both purification and detection.
\item  In contrast to \cite{wu2021adversarial}, which only conducts adversarial defense for the r-vector system, this work performs experiments on both r-vector and x-vector systems to illustrate our proposed defense method is general.
\item  Previous work \cite{wu2021adversarial} only implements one masking strategy to train the self-supervised learning model. This work adopts several masking strategies to train various self-supervised learning models and compare the performance of different masking strategies for adversarial defense. 
\item \cite{wu2021adversarial} only considers the defense against the basic iterative method (BIM), this paper also show the generalizability of our proposed method to other two attack methods, e.g. fast gradient sign method (FGSM) and Jacobian-based saliency map attack (JSMA), without requiring the detailed knowledge of the specific attack methods.
\item  In comparison to \cite{wu2021adversarial}, which simply and crudely uses equal error rate (EER) to evaluate the defense performance, this work systematically formalizes the evaluation metrics for adversarial defense on ASV (see Section~\ref{sec:evaluation-metrics}).
\end{itemize}

\section{Background}
\label{sec:method}

\subsection{Automatic speaker verification}
Automatic speaker verification (ASV) aims at confirming a speaker identity claim given a segment of speech. 
An ASV system usually consists of three parts: feature engineering, speaker embedding extraction, and similarity scoring. 
Feature engineering converts an input waveform into a frame-level acoustic feature sequence, such as Mel-frequency cepstral coefficients (MFCCs), filter-banks, and spectrograms. 
These acoustic features contain variation trends in frequency domain that can reflect speaker cues. 
Speaker embedding extraction aims at projecting acoustic features of an utterance into a speaker-related embedding space. 
It can be conducted by a variety of speaker extraction front-ends, such as i-vector system \cite{dehak2010front,kenny2012small,prince2007probabilistic,garcia2011analysis}, time delay neural network (TDNN) x-vector system \cite{snyder2018x,xu2018generative,li2020bayesian,desplanques2020ecapa} and ResNet r-vector system \cite{zeinali2019but,zhou2021resnext,zhang2017end,seo2019shortcut}. 
Similarity scoring defines a distance metric in the speaker embedding space to measure the speaker similarity between the enrollment and testing embeddings. 
The higher the score is, the more likely the two utterances belong to the same speaker. 
The scoring function usually adopts cosine similarity \cite{dehak2010cosine} and probabilistic linear discriminant analysis (PLDA) \cite{kenny2013plda} functions.

To formulate an ASV system mathematically, we denote the speaker embedding extraction as $g$ with parameters $\boldsymbol{\theta_1}$ and the similarity scoring function as $S$ with parameters $\boldsymbol{\theta_2}$. 
Given an enrollment acoustic feature sequence $\boldsymbol{X_i}$ and a testing acoustic feature sequence $\boldsymbol{X_j}$, the extracted speaker embeddings are derived as
\begin{equation}
    e_i = g_{\boldsymbol{\theta_1}} (\boldsymbol{X_i}), e_j = g_{\boldsymbol{\theta_1}} (\boldsymbol{X_j})
\end{equation}
and the output similarity score $s$ is formulated as
\begin{equation}
    s = S_{\boldsymbol{\theta_2}} (e_i, e_j)
\end{equation}

\subsection{Adversarial sample generation}
As previously discussed, attackers perform adversarial attacks by designing tiny perturbations and adding them to genuine samples to generate adversarial samples.
In a realistic scenario, the enrollment utterance is registered into the ASV system in advance, while the testing utterance is produced by the customer for identity confirmation. 
From adversaries' perspective, they have knowledge about the enrollment utterance, and utilize it along with the ASV function to generate an adversarial testing utterance to fool the ASV.
The ASV system to be attacked is named as target system. If the adversary has the access to the whole parameters of the target system, it can utilize the target system directly to find the adversarial perturbations. Otherwise, the adversary needs to develop his/her own ASV system, and utilizes it as a substitute system to craft adversarial samples. 

There are two kinds of trials in realistic scenarios, i.e. target trials and non-target trials. In target trials, the speaker identities of the enrollment and testing utterances are the same. While in non-target trials, the speaker identities are different. The adversary can perform attacks on both two kinds of trials, as we will show in the following.

Suppose we have an enrollment feature sequence $\boldsymbol{X_{i}}$ and a testing feature sequence $\boldsymbol{X_{j}}$ in each trial. We add perturbations $\boldsymbol{\delta}$ into the testing feature sequence $\boldsymbol{X_j}$ to fool the ASV. This adversarial attack problem for ASV can be formulated as an optimization task, whose objective is to search for a perturbation $\boldsymbol{\delta}$ and add it on the feature sequence $\boldsymbol{X_j}$ to let the system behave incorrectly.
The optimization problem is illustrated as
\begin{equation}
    \boldsymbol{\delta} = \arg \max_{\Vert \boldsymbol{\delta} \Vert_p \leq \epsilon} k \times S_{\boldsymbol{\theta_2}}(g_{\boldsymbol{\theta_1}}(\boldsymbol{X_i}), g_{\boldsymbol{\theta_1}}(\boldsymbol{X_j}+\boldsymbol{\delta})) \label{eq:opt-formulation}
\end{equation}
where the constraint $p$-norm of $\boldsymbol{\delta}$ within perturbation degree $\epsilon$ guarantees that the perturbation is subtle enough so that human cannot perceive the difference between adversarial and genuine samples. $k$ is the trial indicator, derived as
\begin{equation}
    k = \begin{cases}
     -1, & \text{target trial} \\
     1, & \text{non-target trial}
    \end{cases}
    \label{eq:opt-formulation-k}
\end{equation}
For target trials where $k=-1$, attackers aim at enlarging the distance of the enrollment and testing representations and minimizing the similarity score so that the ASV will falsely reject them. While for non-target trials where $k=1$, attackers aim at making the enrollment and testing representations closer to each other, and maximizing the similarity score so that the ASV will falsely accept the imposters.

Different attacking algorithms utilize different adversarial attack strategies.
This work adopts three algorithms to generate adversarial samples, i.e. fast gradient sign method (FGSM)\cite{goodfellow2014explaining}, basic iterative method (BIM)\cite{kurakin2016adversarial_2} and Jacobian-based saliency map attack (JSMA)\cite{papernot2016limitations}. They have been certified to be effective to attack ASV systems\cite{li2020adversarial,li2020investigating}.
We attack the target ASV system directly, i.e. the r-vector (or x-vector) system in this work, to calculate gradients for adversarial perturbation generation, unless otherwise specified.

\subsubsection{Fast gradient sign method (FGSM)}
FGSM specializes the norm $p$ in (\ref{eq:opt-formulation}) as $\infty$ to derive the optimization problem:
\begin{equation}
    \boldsymbol{\delta} = \arg \max_{\Vert \boldsymbol{\delta} \Vert_\infty \leq \epsilon} k \times S_{\boldsymbol{\theta_2}}(g_{\boldsymbol{\theta_1}}(\boldsymbol{X_i}), g_{\boldsymbol{\theta_1}}(\boldsymbol{X_j}+\boldsymbol{\delta})) \label{eq:opt-formulation-fgsm}
\end{equation}
Then, it perturbs the genuine input $\boldsymbol{X_j}$ towards the gradient of the similarity score $S$ w.r.t. $\boldsymbol{X_j}$ multiplied by the trial indicator $k$ to generate adversarial samples. The solutions are given as follows:
\begin{equation}
    \boldsymbol{\delta} = \epsilon \times k \times sign(\nabla_{\boldsymbol{X_j}} S_{\boldsymbol{\theta_2}}(g_{\boldsymbol{\theta_1}}(\boldsymbol{X_i}), g_{\boldsymbol{\theta_1}}(\boldsymbol{X_j}))) \label{eq:fgsm-solu}
\end{equation}
where the function $sign(\cdot)$ takes the sign of the gradient, and $\epsilon$ represents the perturbation degree. The $\epsilon$ is set as 0.3 and 0.6 for attacking the r-vector and x-vector systems, respectively. As we observe that the x-vector system is more hard to attack under the same perturbation degree, we set a larger $\epsilon$ for the x-vector system for a comparable attacking performance.

\subsubsection{Basic iterative method (BIM)}
BIM also specializes the norm $p$ as $\infty$. It perturbs the genuine input $\boldsymbol{X_j}$ in a multi-step manner. Starting from the genuine input $\boldsymbol{X_j^0} = \boldsymbol{X_j}$, the input is perturbed iteratively as
\begin{equation}
    \boldsymbol{\tilde{X}_{j}^{n+1}} = clip_{\boldsymbol{X_{j}}, \epsilon}(\boldsymbol{\tilde{X}_{j}^{n}}+ \boldsymbol{\delta}), ~~~\text{for $n = 0, ..., N-1$} \label{eq:bim-solu-1}
\end{equation}
where $\boldsymbol{\delta}$ is the crafted perturbation, derived as
\begin{equation}
    \boldsymbol{\delta} = \alpha \times sign(\nabla_{\boldsymbol{\tilde{X}_{j}^{n}}} k \times S_{\boldsymbol{\theta_2}}(g_{\boldsymbol{\theta_1}}(\boldsymbol{X_{i}}), g_{\boldsymbol{\theta_1}}(\boldsymbol{\tilde{X}_{j}^{n}})))
    \label{eq:bim-solu-2}
\end{equation}
$sign(\cdot)$ is a function that takes the sign of the gradient, $\alpha$ is the step size, $N$ is the number of iterations and $clip_{\boldsymbol{X_j}, \epsilon}(\boldsymbol{X})$ holds the norm constraints by applying element-wise clipping such that $\Vert \boldsymbol{X}-\boldsymbol{X_j} \Vert_{\infty} \leq \epsilon$. In our experiments, $N$ is set as 5, and $\alpha$ is set as the perturbation degree $\epsilon$ divided by $N$. Similar with the FGSM, the $\epsilon$ is set as 0.3 and 0.6 for attacking the r-vector and x-vector systems, respectively.

As shown in (\ref{eq:bim-solu-1}), BIM perturbs $\boldsymbol{X_{j}}$ towards the gradient of the objective $S$ w.r.t. $\boldsymbol{X_{j}}$ multiplied by the trial indicator $k$ in an iterative manner. Such an iterative optimization strategy can efficiently figure out adversarial perturbations to make ASV behave incorrectly. Finally, the perturbed $\boldsymbol{X_j^{N}}$ is utilized as the crafted adversarial sample.

\begin{algorithm}
\caption{The JSMA perturbation method\\
$\boldsymbol{X_i}$ and $\boldsymbol{X_j}$ are acoustic feature sequences of enrollment and testing utterances, respectively.\\
$k$ is the trial indicator.\\
$g_{\boldsymbol{\theta_1}}$ and $S_{\boldsymbol{\theta_2}}$ represent the speaker embedding extractor and scoring function with parameters, respectively.\\
$\alpha$ is the step size, $\epsilon$ is the perturbation degree, and $N$ is the number of iterations.\\
$\boldsymbol{\Gamma}$ is a mask matrix having the same size with $\boldsymbol{X_j}$, and it masks out the bits that already meet the norm constraints before the bit selection procedure. It is initialized with an all-one element matrix $\boldsymbol{E}$.}
\label{alg:jsma-method}
\begin{algorithmic}[1]
\REQUIRE $\boldsymbol{X_i}$, $\boldsymbol{X_j}$, $k$, $g_{\boldsymbol{\theta_1}}$, $S_{\boldsymbol{\theta_2}}$, $\alpha$, $\epsilon$ and $N$
\STATE $\boldsymbol{X_j^{adv}}=\boldsymbol{X_j}$, $\boldsymbol{\Gamma} = \boldsymbol{E}$, $\boldsymbol{\delta}=\boldsymbol{0}$
\label{code:jsma-2}
\FOR{$i \in [1,N]$}
\label{code:jsma-3}
\STATE $\boldsymbol{G} = \nabla_{\boldsymbol{X_j^{adv}}} k \times S_{\boldsymbol{\theta_2}}(g_{\boldsymbol{\theta_1}}(\boldsymbol{X_i}), g_{\boldsymbol{\theta_1}}(\boldsymbol{X_j^{adv}}))$
\label{code:jsma-4}
\STATE $\boldsymbol{M} = saliency\_map(\boldsymbol{G}, \boldsymbol{\Gamma})$
\label{code:jsma-5}
\STATE $l_{max} = \arg \max_{l} \boldsymbol{M}_l$
\label{code:jsma-6}
\STATE $\boldsymbol{\delta}[l_{max}] = clip_{\boldsymbol{0},\epsilon} (\boldsymbol{\delta}[l_{max}]+\alpha \times sign(\boldsymbol{G}_{l_{max}}))$
\label{code:jsma-7}
\IF{$| \boldsymbol{\delta}[l_{max}] | \geq \epsilon$}
\label{code:jsma-8}
\STATE $\boldsymbol{\Gamma}_{l_{max}} = 0$
\label{code:jsma-9}
\ENDIF
\label{code:jsma-10}
\STATE $\boldsymbol{X_j^{adv}} = \boldsymbol{X_j}+\boldsymbol{\delta}$
\label{code:jsma-11}
\ENDFOR
\label{code:jsma-12}
\RETURN $\boldsymbol{X_j^{adv}}$
\label{code:jsma-13}
\end{algorithmic}
\end{algorithm}

\subsubsection{Jacobian-based saliency map attack (JSMA)}
JSMA is another effective perturbation method to craft adversarial samples. Unlike FGSM and BIM that add perturbations to the whole input, JSMA perturbs only one bit at one iteration. In each iteration, it selects the bit with the most significant effects to achieve successful attack and let it to be perturbed. 
With this purpose, a saliency score is calculated for each bit and the bit with the highest saliency score is chosen to be perturbed. We formulate the algorithm specialized in our case, as shown in Algorithm~\ref{alg:jsma-method}.

The $saliency\_map$ at Step~\ref{code:jsma-5} computes the absolute value of gradient $\boldsymbol{G}$ while masking out the bits already reach the constraint boundary: $saliency\_map(\boldsymbol{G}, \boldsymbol{\Gamma})= \boldsymbol{G}^{abs} \odot \boldsymbol{\Gamma}$, where $\boldsymbol{G}^{abs}$ is the element-wise absolute value of $\boldsymbol{G}$ and $\odot$ is an element-wise product operator. In this work, $N$ is set as 300 iterations, and $\alpha$ is set as half of the perturbation degree $\epsilon$. The $\epsilon$ is set as 2.0 for both the r-vector and x-vector systems. Finally, $\boldsymbol{X_j^{adv}}$ is utilized as the designed adversarial sample.

\section{Self-supervised learning}
\label{sec:Self-supervised learning}
Self-supervised learning has been widely applied in different domains; here, we mainly focus on self-supervised learning for speech processing.
There are two major branches of self-supervised learning methods in speech domain: contrastive predictive coding (CPC) and reconstruction.

The CPC method~\cite{cpc, modified_cpc, bidir_cpc, wav2vec, vq_wav2vec, vq_wav2vec_ft} uses auto-regressive modeling of representations in the feature space, where the model learns to determine the near future acoustic frames while contrasting with frames from a more distant temporal location.
The contrastive loss pulls temporally nearby representations closer and pushes temporally distant ones further.
The CPC contrastive objective does not directly predict acoustic features, so it is not suitable for this work.

On the other hand, another branch of work uses reconstruction objectives for self-supervised learning.
Unlike the CPC objective, acoustic features are predicted as the models' output, and a reconstruction loss is computed across the predicted features and ground-truth features.
The auto-regressive predictive coding (APC) methods ~\cite{apc1, apc2} predict future frames like a recurrent-based language model (LM)~\cite{rnn} conditioning on past frames.
In other approaches~\cite{tera, mockingjay, audioalbert}, parts of the input acoustic features are randomly masked by a variety of masking strategies, and the L1 reconstruction loss is set as the learning objective.
The random masking strategies of these approaches ~\cite{tera, mockingjay, audioalbert} are inspired by the masked language model (MLM) task from BERT~\cite{bert, roberta, albert}; hence these models can be seen as a speech version of BERT.
Their inference time is fast since they are Transformer~\cite{transformer} models which can process the input sequence in parallel (unlike auto-regressive models).

In this work, we employ the speech BERT trained by denoising tasks as self-supervised learning based reformer to do defense.
Inspired by~\cite{tera, mockingjay, audioalbert, naseer2020self, chen2020adversarial, deb2020faceguard}, we adopt a variety of masking strategies to perturb the inputs and train the self-supervised learning models. During training, the perturbed acoustic features are fed into the SSLR models, and the L1 reconstruction loss between the outputs and the clean acoustic features is derived as the loss function to optimize model parameters.
Fig.~\ref{fig:mfcc} illustrates a sample of 24-dimensional MFCC feature sequence from the Voxceleb1 dataset.
Fig.~\ref{fig:mfcc}.(a) is the original plot of MFCC, and three masking strategies are shown below:
\begin{figure}[ht]
\centering

\subfigure[]{
\includegraphics[width=4.0cm]{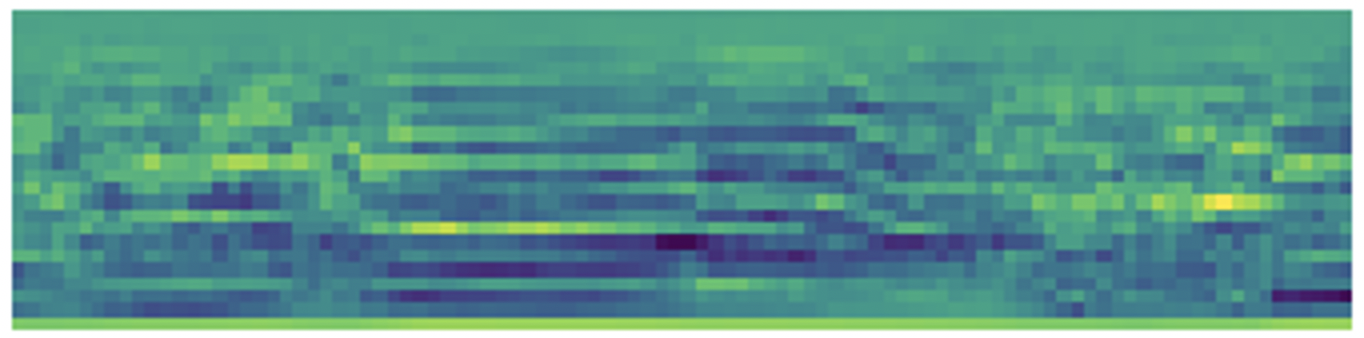}

}
\subfigure[]{
\includegraphics[width=4.0cm]{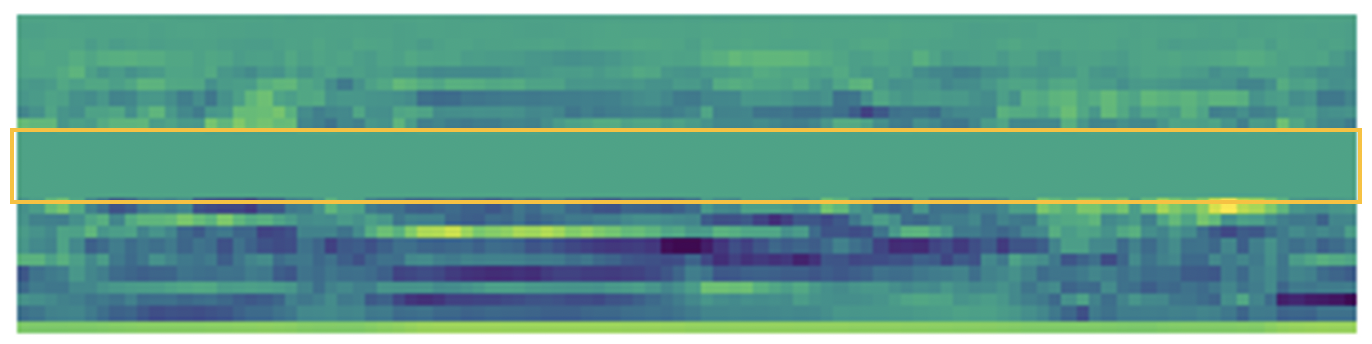}
}
\quad
\subfigure[]{
\includegraphics[width=4.0cm]{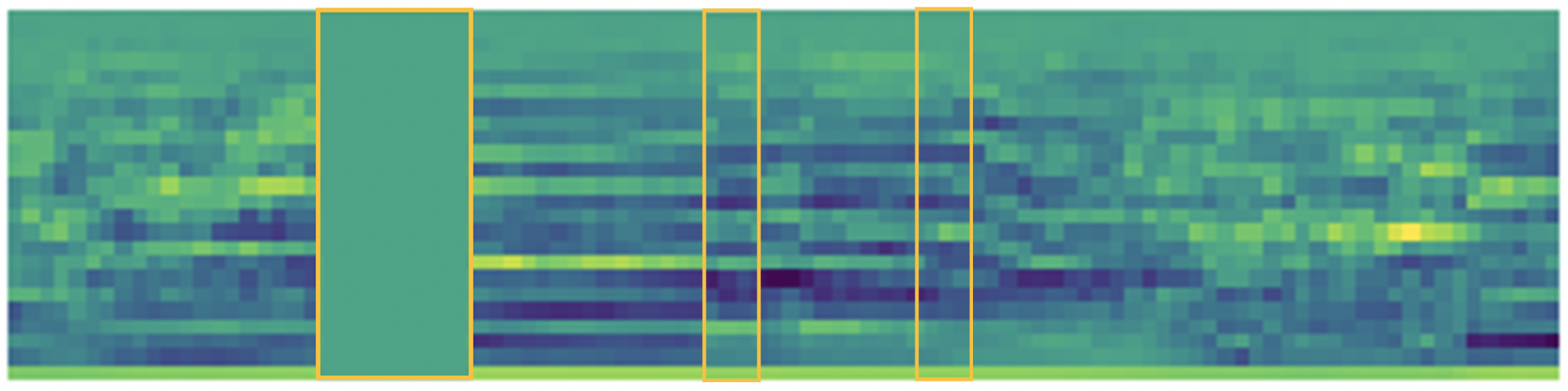}
}
\subfigure[]{
\includegraphics[width=4.0cm]{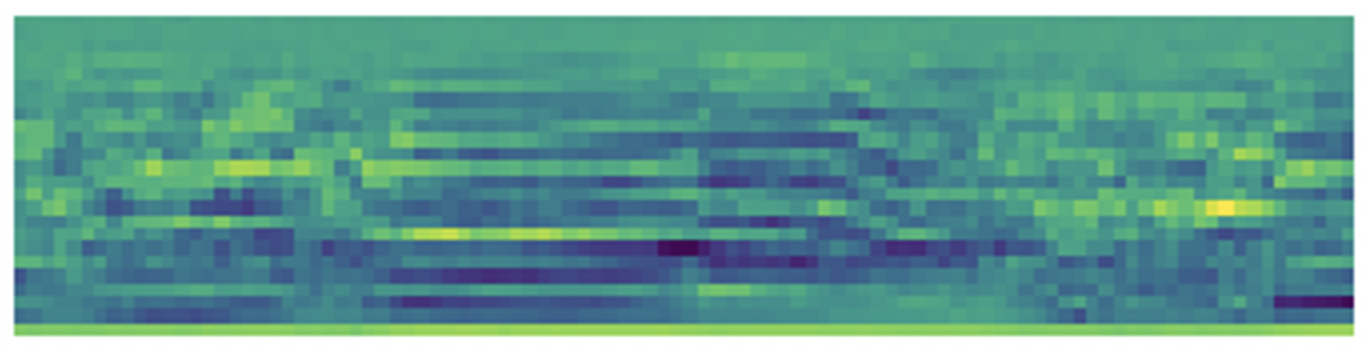}

}

\caption{The illustration of inputs with various masking strategies. The masked part is highlighted by an orange block. (a) is the original MFCC, and (b), (c), (d) are the MFCC modified by channel masking, time masking and magnitude alteration, respectively.}
\label{fig:mfcc}
\end{figure}
\begin{itemize}
    \item In Fig.~\ref{fig:mfcc}.(b), we adopt the channel masking by randomly selecting a block of consecutive channels with length $W_{C}$ and masking the values within the block to zero for all time steps across the input sequence. The channel masking strategy is originated from SpecAugment \cite{park2019specaugment}. During training, the masked features are fed into the SSLR. However, clean samples may be taken as inputs into the SSLR models during the inference stage, which will cause inconsistency and mismatch problems. So, for $1/(W_{C} +1)$ of the time during training, none of the channels will be masked such that the model can receive the entire channel information, which will alleviate the mismatch problem of model's inputs between training time and inference time.
    \item Fig.~\ref{fig:mfcc}.(c) illustrates the time masking process. We randomly select $P_T$ percentage frames out of the total frames, and the selected $P_T$ percentage frames are made of several blocks of contiguous frames with width $W_{T}$ along time. Then, we do masking within the frames according to the following strategy: 1) At a probability of 0.8, all the selected frames are set to zero. (the first orange block in Fig.~\ref{fig:mfcc}.(c)). 2) At a probability of 0.1, we randomly select other frames to replace them. (the second orange block in Fig.~\ref{fig:mfcc}.(c)). 3) At a probability of 0.1, the selected frames remain unchanged. (the third orange block in Fig.~\ref{fig:mfcc}.(c)). Case 3) also addresses the inconsistency problem between training and inference. Through time masking, the model needs to do reconstruction by bidirectional representations which contain both past and future context. 
    \item We do magnitude alteration by adding Gaussian noise to the whole input acoustic feature sequence with probability $P_{N}$, as shown in Fig.~\ref{fig:mfcc}.(d). Magnitude alteration can be seen as a kind of data augmentation.
\end{itemize}
The self-supervised learning based reformer trained by channel masking, time masking, and magnitude alteration are denoted as SSLR-C, SSLR-T, and SSLR-M, respectively.
The permutation of the above three masking methods is adopted to enlarge and diversify the pool of SSLR models, i.e. SSLR-TC\footnote{SSLR-TC means the self-supervised learning reformer trained by the combination of time masking and channel masking.}, SSLR-TM, SSLR-CM and SSLR-TCM.

\label{sec:method}

\begin{figure*}[ht]
  \centering
  \centerline{\includegraphics[width=\linewidth]{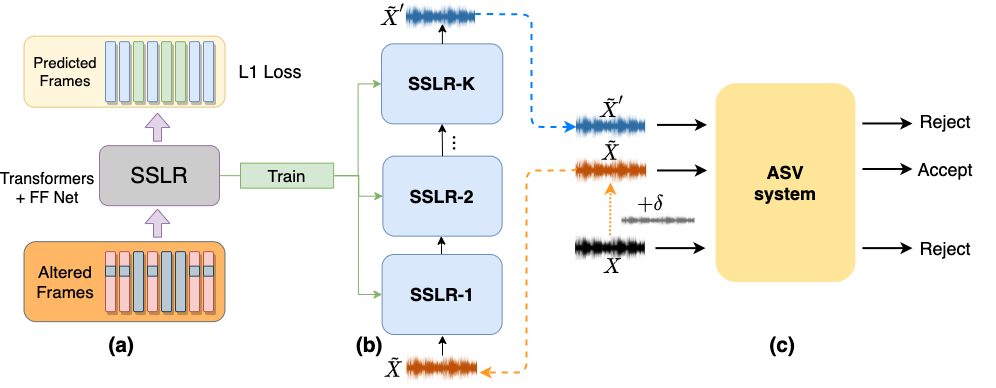}}
  \caption{(a) Illustration of the SSLR models training. The gray pixels in the altered frames (the orange block in (a) ) means masking. The green and blue frames in the predicted frames (the yellow block in (a) ), are the reconstructed frames for time masking and frequency masking respectively. (b) Adversarial defense by cascaded SSLR models. The cascaded SSLR models in (b) are only used during inference. (c) Automatic speaker verification. Notice that all $\boldsymbol{X}$, $\boldsymbol{\tilde{X}}$, $\boldsymbol{\tilde{X}'}$ and $\boldsymbol{\delta}$ in the figure are matrices that represent acoustic feature sequences.}
  \label{fig:defense}
\end{figure*}

\section{Proposed adversarial defense method}
We adopt the self-supervised learning based reformer (SSLR) for adversarial defense as it attains the ability of purifying the superficial perturbations and maintaining the pivotal information in the inputs after training.
Based on the SSLR, we propose a defense framework which includes one purification module, named adversarial perturbation purification, and one detection module, named adversarial perturbation detection, for defending ASV systems against adversarial samples. 
In contrast to existing works \cite{wang2019adversarial,li2020investigating,zhang2020adversarial}, our proposed framework doesn't require knowledge of the process for adversarial sample synthesis.
The two modules in our defense framework are orthogonal to adversarial training \cite{wang2019adversarial,wu2020defense}.
The purification module purifies tiny adversarial perturbations and recovers the clean samples from adversarial ones.
The detection module aims at differentiating the adversarial samples from genuine samples by approximating the statistical properties of genuine samples and setting a threshold.

\subsection{Adversarial perturbation purification}
As adversarial perturbation is also a kind of noise to some extent, we firstly train the SSLR models with adequate data to learn the ability of purifying adversarial noise during the reconstruction process.
Note that, we use the masking strategies to train SSLR models and then fix the parameters. We won't leverage adversarial samples to fine-tune the SSLM models.
From the perspective of attackers, they attempt to manipulate the ASV systems into predicting scores higher than the threshold for non-target trials, and scores lower than the threshold for target trials by crafting adversarial perturbations and adding them to the genuine inputs.
We use our proposed defense module, cascaded SSLR models, to purify the adversarial noise, as shown in Fig.~\ref{fig:defense}.b.
Each SSLR slot in Fig.~\ref{fig:defense}.b will do reconstruction of the inputs and help alleviate the superficial noise in the inputs.
Different masking configurations will equip SSLR models with different degrees of abilities for denoising.
In order to enlarge the diversity of self-supervised learning models and compare the denoising abilities of different masking configurations, we trained $M$ SSLR models with different masking settings in advance.
As shown in Fig.~\ref{fig:defense}.b, we select one of the $M$ SSLR models, and duplicate it into K models to be put into the $K$ SSLR blocks. This means the SSLR models put into $K$ blocks are identical.
During inference, before the audio is fed into the ASV system, the purification module will alleviate the adversarial noise block by block and recover the clean counterpart.



\begin{figure*}[ht]
  \centering
  \centerline{\includegraphics[width=\linewidth]{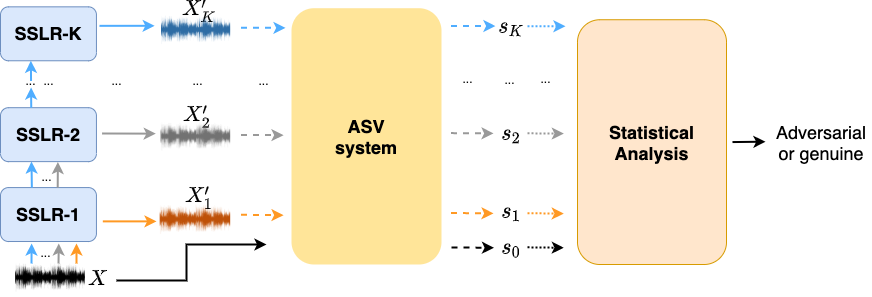}}
  \caption{Flow of detection. All $\boldsymbol{X}$ and \{$\boldsymbol{\tilde{X}_i} \vert i = 1, ..., K$\} in the figure are matrices that represent acoustic feature sequences of an utterance.}
  \label{fig:detection}
\end{figure*}

\subsection{Adversarial perturbation detection}
The SSLR models attain the capacity to purify the superficial noise in the adversarial samples and pull them back towards the decision boundary, while keep the key information of genuine samples and do nearly lossless reconstruction.
Take a non-target trial as an example, the score of the genuine sample is below the threshold, and the ASV scores change little after being transformed by the cascaded SSLR models. 
As a result, the variation of ASV scores of genuine samples after passing through different numbers of SSLR models is small.
By contrast, the ASV score for the adversarial sample is over the threshold.
The SSLR models will try to purify the adversarial noise and decrease the ASV score of the adversarial sample to make the score below the threshold.
So the variations for adversarial samples will change more significantly than genuine samples after being transformed by the cascaded SSLR models.
Based on it, we want to harness the variation of ASV scores to distinguish adversarial samples from genuine samples.
Specifically, as illustrated in Fig.~\ref{fig:detection}, we concatenate $K$ SSLR blocks and put one SSLR model into each block.
Then we provide the input $\boldsymbol{X}$ into SSLR-1, passing it through different systems consisting of different numbers of cascaded SSLR models to generate reconstructed outputs $\{\boldsymbol{X'_{1}}, \boldsymbol{X'_{2}} ..., \boldsymbol{X'_{K}}\}$.
Based on $\{\boldsymbol{X}, \boldsymbol{X'_{1}}, \boldsymbol{X'_{2}} ..., \boldsymbol{X'_{K}}\}$, $K+1$ ASV scores $\{s'_{0}, s'_{1}, s'_{2} ..., s'_{K}\}$ can be derived, as illustrated in Fig.~\ref{fig:detection}.
We notice that the score variations of $\{s'_{0}, s'_{1}, s'_{2} ..., s'_{K}\}$ are different for genuine and adversarial samples, i.e. adversarial samples have larger score variations than genuine ones. Thereby, we decide to apply statistical moments to illustrate the score variations and use them to distinguish adversarial and genuine samples. Specifically, we suppose that we derive $N$ scores for a trial, denoted by $s_n$ where $n \in \{1, 2, ..., N\}$. Then, we compute the $k$-th moment of these scores as
\begin{equation}
    m_k = \frac{1}{N} \sum_{n=1}^{N} (s_n - \overline{s})^k \label{eq:moment-2}
\end{equation}
where $\overline{s}$ is the mean of these ASV scores:
\begin{equation}
    \overline{s} = \frac{1}{N} \sum_{n=1}^{N} s_n \label{eq:moment-1}
\end{equation}
Since the sign of each moment has no information on the scale of the variation, we just use the absolute value of the moment ($t_k$) to indicate whether a sample is adversarial or not as
\begin{equation}
    t_k = \vert m_k \vert \label{eq:moment-3}
\end{equation}
Finally, $t_k$ is utilized as the detection score $d_i$ for each trial $i$.




\subsection{Evaluation Metrics}
\label{sec:evaluation-metrics}

\begin{table*}[h]
\caption{The trial subset of each evaluation metric.}
\label{tab:eval-metric-comparison}
\centering
\begin{tabular}{c|c|c|c|c}
\hline
\hline
          & GenFAR & AdvFAR & j-FAR\_pur & j-FAR\_det \\
\hline
Trial subset & $\mathbb{T}_{gen\_ntgt}$ & $\mathbb{T}_{adv\_ntgt}$ & $\mathbb{T}_{gen\_ntgt} \cup \mathbb{T}_{adv\_ntgt}$ & $\mathbb{T}_{gen\_ntgt} \cup \mathbb{T}_{adv\_ntgt} \cup \mathbb{T}_{adv\_tgt}$ \\
\hline
\hline
          & GenFRR & AdvFRR & j-FRR\_pur & j-FRR\_det \\
\hline
Trial subset & $\mathbb{T}_{gen\_tgt}$ & $\mathbb{T}_{adv\_tgt}$ & $\mathbb{T}_{gen\_tgt} \cup \mathbb{T}_{adv\_tgt}$ & $\mathbb{T}_{gen\_tgt}$ \\
\hline
\hline
\end{tabular}
\end{table*}

Since previous works on ASV adversarial defense are rare and limited, and also there is not a common metric to evaluate the defense performance, we propose several evaluation metrics in this work.
Before that, we introduce some common evaluation metrics for ASV, i.e. false acceptance rate (FAR), false rejection rate (FRR) and equal error rate (EER). FAR represents the percentage of non-target trials in which unauthorised persons are incorrectly accepted. Accordingly, FRR is the percentage of target trials where authorised persons are incorrectly rejected. Finally, EER is the error rate under the operation point where FAR equals to FRR.

For adversarial perturbation purification module, we propose to utilize GenEER to evaluate the performance of ASV systems on genuine samples, while utilize AdvFAR and AdvFRR to evaluate the performance of ASV under adversarial attacks. In realistic scenarios, system designers are usually not aware of adversarial samples and determine the system operation point $\tau_{ASV}$ that minimizes certain metrics evaluated on genuine samples only. 
To this end, we determine the $\tau_{ASV}$ based on the EER metrics on genuine samples:
\begin{equation}
    \tau_{ASV} = \{ \tau \in \mathbb{R} : GenFAR(\tau) = GenFRR(\tau) \} \label{eq:asv-threshold}
\end{equation}
We have defined GenFAR and GenFRR as
\begin{equation}
    GenFAR(\tau) = \frac{\vert \{s_i \geq \tau : i \in \mathbb{T}_{gen\_ntgt}\} \vert}{\vert \mathbb{T}_{gen\_ntgt} \vert} \label{eq:gen-far}
\end{equation}
and
\begin{equation}
    GenFRR(\tau) = \frac{\vert \{s_i < \tau : i \in \mathbb{T}_{gen\_tgt}\} \vert}{\vert \mathbb{T}_{gen\_tgt} \vert} \label{eq:gen-frr}
\end{equation}
respectively, where $\mathbb{T}_{gen\_tgt}$ and $\mathbb{T}_{gen\_ntgt}$ represent the trial sets consisting of genuine target and genuine non-target trials, respectively. $s_i$ denotes the ASV score for the $i^{th}$ trial and $\vert \mathbb{A} \vert$ denotes the number of elements in set $\mathbb{A}$. Finally, GenEER that reflects the system EER on genuine samples is derived as
\begin{equation}
    GenEER = GenFAR(\tau_{ASV}) = GenFRR(\tau_{ASV}) \label{eq:gen-eer}
\end{equation}
To evaluate ASV performance on adversarial samples, given the system operation point $\tau_{ASV}$, AdvFAR and AdvFRR are derived as
\begin{equation}
    AdvFAR = \frac{\vert \{s_i \geq \tau_{ASV} : i \in \mathbb{T}_{adv\_ntgt} \} \vert}{\vert \mathbb{T}_{adv\_ntgt} \vert} \label{eq:adv-far}
\end{equation}
and
\begin{equation}
    AdvFRR = \frac{\vert \{ s_i < \tau_{ASV} : i \in \mathbb{T}_{adv\_tgt} \} \vert}{\vert \mathbb{T}_{adv\_tgt} \vert} \label{eq:adv-frr}
\end{equation}
respectively, where $\mathbb{T}_{adv\_tgt}$ and $\mathbb{T}_{adv\_ntgt}$ represent the trial sets consisting of adversarial target and adversarial non-target trials, respectively. Finally, we evaluate our adversarial purification module in two aspects: 1) the reduction of system's AdvFAR and AdvFRR under adversarial attacks; 2) the influence on system's GenEER on genuine samples.

For adversarial perturbation detection module, we propose to utilize $EER_{det}$ for evaluating the detector's discrimination ability between adversarial and genuine samples. The $EER_{det}$ is derived as
\begin{equation}
    EER_{det} = FAR_{det}(\tau_{det}) = FRR_{det}(\tau_{det}) \label{eq:det-eer}
\end{equation}
where $\tau_{det}$ is a determined detection threshold that makes the FAR of the adversarial samples and the FRR of the genuine samples equal to each other:
\begin{equation}
    \tau_{det} = \{ \tau \in \mathbb{R} : FAR_{det}(\tau) = FRR_{det}(\tau) \} \label{eq:det-threshold}
\end{equation}
We have defined $FAR_{det}$ and $FRR_{det}$ as
\begin{equation}
    FAR_{det}(\tau) = \frac{\vert \{ d_i < \tau : i \in \mathbb{T}_{adv} \} \vert}{\vert \mathbb{T}_{adv} \vert} \label{eq:det-far}
\end{equation}
and
\begin{equation}
    FRR_{det}(\tau) = \frac{\vert \{ d_i \geq \tau : i \in \mathbb{T}_{gen} \} \vert}{\vert \mathbb{T}_{gen} \vert} \label{eq:det-frr}
\end{equation}
respectively, where $\mathbb{T}_{adv}$ and $\mathbb{T}_{gen}$ denote the trial sets consisting of adversarial and genuine trials, respectively.

Moreover, with the consideration that adversarial and genuine samples are pooled together in realistic scenarios, we further pool adversarial and genuine samples together and compute a joint-FAR (j-FAR) and a joint-FRR (j-FRR) for system evaluation. These two metrics can reflect a trade-off performance of our defense approaches between alleviating adversarial attacks and preserving ASV performance on genuine samples. Our experiments are conducted where adversarial and genuine samples are combined in equal proportion, unless otherwise specified. Moreover, in order to investigate the variation of j-FAR and j-FRR with respective to different proportions of adversarial samples, we also conduct experiments where adversarial and genuine samples are combined with a range of proportion values, as discussed in Section~\ref{sec:prf-unaware-sslr}.

Since the j-FAR and j-FRR formulas are slightly different between the purification and detection modules, we utilize subscripts to distinguish their derivations, i.e. j-FAR\_{pur} and j-FRR\_{pur} for the purification module and j-FAR\_{det} and j-FRR\_{det} for the detection module. These subscripts only appear in the equations, and we still use j-FAR and j-FRR elsewhere for simplicity.

For the purification module, we pool adversarial and genuine samples together with a certain portion, then derive j-FAR\_{pur} and j-FRR\_{pur} as
\begin{equation}
    \text{j-FAR\_{pur}} = \frac{\vert \{ s_i \geq \tau_{ASV} : i \in \mathbb{T}_{pur\_joint\_ntgt} \} \vert}{\vert \mathbb{T}_{pur\_joint\_ntgt} \vert} \label{eq:prf-jfar}
\end{equation}
and
\begin{equation}
    \text{j-FRR\_{pur}} = \frac{\vert \{ s_i < \tau_{ASV} : i \in \mathbb{T}_{pur\_joint\_tgt} \} \vert}{\vert \mathbb{T}_{pur\_joint\_tgt} \vert} \label{eq:prf-jfrr}
\end{equation}
respectively. Here, we have defined
\begin{equation}
    \mathbb{T}_{pur\_joint\_ntgt} = \mathbb{T}_{adv\_ntgt} \cup \mathbb{T}_{gen\_ntgt} \label{eq:prf-ntgt-set}
\end{equation}
and
\begin{equation}
    \mathbb{T}_{pur\_joint\_tgt} = \mathbb{T}_{adv\_tgt} \cup \mathbb{T}_{gen\_tgt} \label{eq:prf-tgt-set}
\end{equation}

While for the detection module, notice that the detector aims at detecting all adversarial samples and rejecting them directly, the ground truth operation for all adversarial trials is rejection. Hence, different from the trial set partition for the purification module (in (\ref{eq:prf-ntgt-set}) and (\ref{eq:prf-tgt-set})), we categorize the adversarial target trials into the partition of non-target trials for the detection module. The partition of target and non-target trials for detection are shown as
\begin{equation}
    \mathbb{T}_{det\_joint\_tgt} =  \mathbb{T}_{gen\_tgt} \label{eq:det-tgt-set}
\end{equation}
and
\begin{equation}
    \mathbb{T}_{det\_joint\_ntgt} = \mathbb{T}_{adv\_ntgt} \cup \mathbb{T}_{adv\_tgt} \cup \mathbb{T}_{gen\_ntgt} \label{eq:det-ntgt-set}
\end{equation}
respectively. Then, given the operation points of both the detector and ASV, j-FAR\_{det} and j-FRR\_{det} are derived as
\begin{equation}
    \text{j-FAR\_{det}} = \frac{\vert \{ d_i < \tau_{det} \cap s_i >= \tau_{ASV} : i \in \mathbb{T}_{det\_joint\_ntgt} \} \vert}{\vert \mathbb{T}_{det\_joint\_ntgt} \vert} \label{eq:det-jfar}
\end{equation}
and
\begin{equation}
    \text{j-FRR\_{det}} = \frac{\vert \{ d_i \geq \tau_{det} \cup s_i < \tau_{ASV} : i \in \mathbb{T}_{det\_joint\_tgt} \} \vert}{\vert \mathbb{T}_{det\_joint\_tgt} \vert}  \label{eq:det-jfrr}
\end{equation}
respectively.

Notice that apart from GenEER that performs the EER computation on genuine target and non-target trials, other performance metrics are based on the FAR/FRR computation by just operating on different trial subsets. Table~\ref{tab:eval-metric-comparison} summarizes the trial subset of each performance metric.




\section{Experimental setup}
\subsection{Dataset}
In this work, we conduct experiments on Voxceleb1 \cite{nagrani2017voxceleb} dataset, which consists of short clips of human speech. There are in total 148,642 utterances for 1251 speakers. We develop our ASV system on the train and development partitions while reserve 4,874 utterances of the testing partition for evaluating our ASV system and generating adversarial samples.
Notice that generating adversarial samples is time-consuming and resource-consuming. Without loss of generality, we randomly select 1000 trials out of 37,720 trials provided in \cite{nagrani2017voxceleb}, to generate adversarial samples.
The randomly selected 1000 trials include 500 non-target trials and 500 target trials.

\subsection{ASV setup}
In this work, we develop two ASV systems to demonstrate the robustness of our proposed methods. They are the x-vector system \cite{snyder2018x} and the r-vector system \cite{zeinali2019but}.

\begin{table*}[th]
\caption{Purification performance consistency between complete and 1K trials.}
\label{tab:purification-performance-consistency}
\centering
\begin{tabular}{c|c|c|c|c|c|c}
\hline
\hline
 \multirow{2}{*}{x-vector system}& \multicolumn{3}{c}{Complete trials} & \multicolumn{3}{|c}{1K trials} \\
 \cline{2-7}
            & AdvFAR (\%) & AdvFRR (\%)  & GenEER (\%)   & AdvFAR (\%) & AdvFRR (\%)  & GenEER (\%) \\
\hline
NA          & 66.21       & 78.98        &  5.93       & 67.29     & 82.25        & 6.59 \\
9*SSLR-TCM  & 26.17       & 24.28        &  10.52        & 26.21       & 27.06        & 11.30 \\
\hline
\hline
 \multirow{2}{*}{r-vector system}& \multicolumn{3}{c}{Complete trials} & \multicolumn{3}{|c}{1K trials} \\
 \cline{2-7}
            & AdvFAR (\%) & AdvFRR (\%)  & GenEER (\%)   & AdvFAR (\%) & AdvFRR (\%)  & GenEER (\%) \\
\hline
NA          & 76.44       & 59.20        &  8.39         &  72.30      & 64.29    &   8.90 \\
9*SSLR-TCM  & 21.68       & 19.63        &  13.50        &  20.82      & 22.73        &   14.59 \\
\hline
\end{tabular}
\end{table*}

The x-vector system is configured as \cite{snyder2018x}, except that additive angular margin (AAM)-softmax loss \cite{xiang2019margin} with hyper-parameters \{$m=0.3$, $s=32$\} is used for training. Extracted x-vectors are centered and projected to a 200-dimensional vector by LDA, then length-normalized before PLDA modeling.

\begin{table}[th]
\caption{ASV performance with genuine and adversarial inputs.}
\label{tab:asv-performance}
\centering
\begin{tabular}{c|c|c|c|c}
\hline
\hline
 \multirow{2}{*}{x-vector system}& \multicolumn{2}{c}{Complete trials} & \multicolumn{2}{|c}{1K trials} \\
 \cline{2-5}
            & EER (\%) & minDCF  & EER (\%)    & minDCF \\
\hline
genuine input          & 5.93    & 0.513     & 6.71 & 0.608 \\
BIM attack             & 72.59   & 1.000     & 73.38  & 1.000 \\
FGSM attack            & 66.83   & 0.9995    & 68.83  & 1.000 \\
JSMA attack            & 56.04   & 0.9948    & 57.36  & 0.9978 \\
\hline
\hline
 \multirow{2}{*}{r-vector system}& \multicolumn{2}{c}{Complete trials} & \multicolumn{2}{|c}{1K trials} \\
 \cline{2-5}
            & EER (\%) & minDCF  & EER (\%)    & minDCF \\
\hline
genuine input          & 8.39    & 0.639       & 8.87  & 0.792 \\
BIM attack             & 65.89   & 1.000     & 66.23 & 1.000 \\
FGSM attack            & 57.08   & 0.9994    & 57.36 & 1.000 \\
JSMA attack            & 79.41   & 0.9997    & 80.52 & 1.000 \\
\hline
\end{tabular}
\end{table}

The r-vector system has the same architecture as \cite{zeinali2019but}, and AAM-softmax loss \cite{xiang2019margin} with hyper-parameters \{$m=0.2$, $s=30$\} is used for training. Extracted r-vectors are centered and length-normalized before cosine scoring.

\subsection{ASV performance with genuine and adversarial inputs}
This section demonstrates the ASV system performance with genuine and adversarial inputs, as well as the performance consistency between the complete trials and the randomly selected 1K trials.

Table~\ref{tab:asv-performance} shows the x-vector and r-vector system performance before and under adversarial attacks on both the complete trials and the randomly selected 1K trials. The system performance is evaluated by equal error rate (EER) and minimum detection cost function (minDCF) with a prior probability of target trials as 0.01. As shown in Table~\ref{tab:asv-performance}, all three adversarial attacks, i.e. BIM, FGSM and JSMA attacks, can seriously degrade both x-vector and r-vector system performance, which verifies the effectiveness of the performed adversarial attacks and the vulnerability of SOTA ASV models to adversarial attacks. Moreover, a performance consistency is observed between the complete trials and the randomly selected 1K trials, which demonstrates that it is reasonable to conduct experiments on the selected 1K trials.

\begin{table}[th]
\caption{Detection performance ($EER_{det} (\%)$) consistency between complete and 1K trials.}
\label{tab:detection-performance-consistency}
\centering
\begin{tabular}{c|c|c|c|c}
\hline
\hline
  Complete trials & 2nd   & 3rd   & 4th   & 5th \\
\hline
  r-vec system    & 26.7  & 21.4  & 24.1  & 21.7 \\
  x-vec system    & 20.7  & 16.6  & 18.5  & 16.6 \\
\hline
\hline
  1K trials       & 2nd   & 3rd   & 4th   & 5th \\
\hline
  r-vec system    & 25.7  & 21.0  & 23.3  & 21.2 \\
  x-vec system    & 21.2  & 16.7  & 19.0  & 16.9 \\
\hline
\hline
\end{tabular}
\end{table}

In order to demonstrate that the proposed defense methods also have a consistent effectiveness between the complete trials and the selected 1K trials, we show the consistency of the purification and detection performance in advance in Table~\ref{tab:purification-performance-consistency} and Table~\ref{tab:detection-performance-consistency}, respectively. Due to space limitation, we only show the results under the BIM attack, and the FGSM and JSMA results have similar trends. In Table~\ref{tab:purification-performance-consistency}, we observe that both the x-vector and r-vector system have very similar performance between the complete trials and the selected 1K trials, either before (``NA'' in the table) or after (``9*SSLR-TCM'' in the table) integrating the SSLR models. In Table~\ref{tab:detection-performance-consistency}, the detection performance ($EER_{det}$) is consistent between the complete and 1K trials as well. The experimental details will be discussed in Section~\ref{sec:prf-all} and Section~\ref{sec:det-all} for the purification and detection methods, respectively.

From the observations above, we conduct further experiments on the selected 1K trials only to save efforts. If not specified, we only show the BIM results for the following experiments due to limited space, and the FGSM and JSMA results have similar trends.

\subsection{SSLR setup}
The training dataset for SSLR models is Voxceleb2 \cite{chung2018Voxceleb2} which includes 2442 hours raw speech.
There are in total 1,128,242 utterances from 6112 speakers in Voxceleb2.
We use both the training and testing set of Voxceleb2 for training the SSLR models.
Seven SSLR models with different masking strategies are trained: SSLR-T, SSLR-M, SSLR-C, SSLR-TM, SSLR-TC, SSLR-CM, SSLR-TCM.
The structures of all SSLR models are identical, and they consist of three layers of transformer encoders with multi-head self attention \cite{vaswani2017attention} and two feed-forward network.
The time masking percentage $P_T$ is 15\%, the width of time masking $W_T$ is 7, the length of channel masking $W_C$ is 5, magnitude alteration probability $P_N$ is set to 0.15, and the Gaussian noise for magnitude alteration is with zero mean and 0.2 variance.
The acoustic feature for SSLR models is 24 dimensional MFCCs, extracted by a pre-emphasis with coefficient of 0.97 and a "Hamming" window having size of 25ms and step-size of 10ms. 
The Adam optimizer \cite{kingma2014adam} is adopted to train the SSLR models guided by L1 reconstruction loss.

\begin{figure*}[h]
    \centering
    \begin{minipage}[h]{0.45\textwidth}
        \centering
        \includegraphics[width=3.5in]{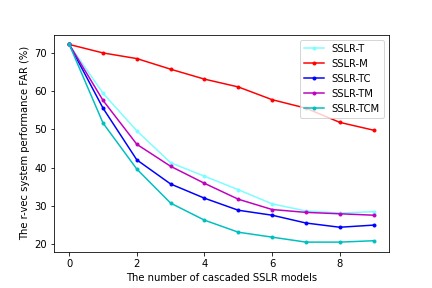}
        {\par \footnotesize (a)\par}
    \end{minipage}
    \hfill
    \begin{minipage}[h]{0.45\textwidth}
        \centering
        \includegraphics[width=3.5in]{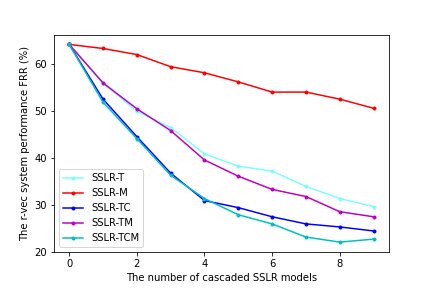}
        {\par \footnotesize (b)\par}
    \end{minipage}
    \hfill
    \begin{minipage}[h]{0.45\textwidth}
        \centering
        \includegraphics[width=3.5in]{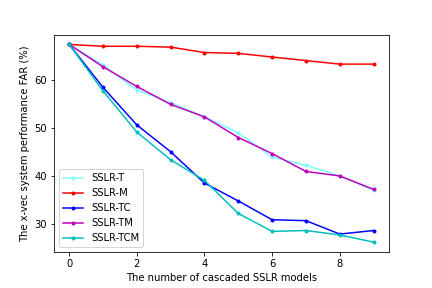}
        {\par \footnotesize (c)\par}
    \end{minipage}
    \hfill
    \begin{minipage}[h]{0.45\textwidth}
        \centering
        \includegraphics[width=3.5in]{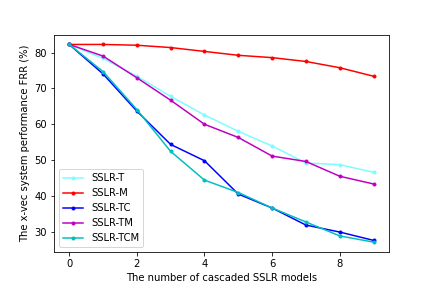}
        {\par \footnotesize (d)\par}
    \end{minipage}
    \caption{ASV system performance with different numbers of cascaded SSLR models. (a) and (b) show the AdvFAR and AdvFRR of the r-vector system, respectively. (c) and (d) show the AdvFAR and AdvFRR of the x-vector system, respectively.}
    \label{fig:purification-unaware-SSLR}
\end{figure*}

\section{Experimental Results: Adversarial Samples Purification}
\label{sec:prf-all}

This section evaluates the adversarial purification performance of the SSLR models. Two situations are taken into account according to the knowledge that attackers know about the SSLR models. Firstly, we assume that attackers are unaware of the SSLR models, and adversarial samples are generated with the ASV model only. This situation is the most practical one since the realistic attacking scenarios are usually black-box ones, and the attackers cannot access the model internals, not to mention the defense method. Secondly, we leave some knowledge of SSLR models to attackers to demonstrate our methods' robustness. 

\subsection{Attackers are unaware of the SSLR models}
\label{sec:prf-unaware-sslr}
In this section, we assume that attackers are unaware of the SSLR models in front of ASV. We mainly show the results of BIM due to space limitation.

The AdvFAR and AdvFRR of the r-vector system integrated with different numbers of SSLR models are shown in Fig.~\ref{fig:purification-unaware-SSLR} (a) and Fig.~\ref{fig:purification-unaware-SSLR} (b), respectively. We observe that AdvFAR and AdvFRR decrease dramatically as ASV integrates more SSLR models, which indicates the effectiveness of our proposed SSLR models on purifying adversarial samples. Specifically, we observe that SSLR models trained by more than one masking strategy have better performance than those with only one masking strategy. Accordingly, SSLR-TCM achieves the best defense performance where AdvFAR decreases from over 75\% to around 20\% and AdvFRR decreases from over 65\% to around 20\%, as shown in Fig.~\ref{fig:purification-unaware-SSLR}. One possible explanation is that integrating different masking strategies strengthens the denoising ability of the SSLR, and then equips the SSLR models with better capacity of purifying adversarial samples. Similar results on the x-vector system are observed in Fig.~\ref{fig:purification-unaware-SSLR} (c) and Fig.~\ref{fig:purification-unaware-SSLR} (d), which indicates that the defense capacity of SSLR models can generalize to different ASV systems.


Because this paper makes the first attempt to conduct adversarial defense by adversarial perturbation purification without knowing the adversarial sample generation procedure and there is no baseline for comparison, we employ traditional filters for defense and set them as baselines.
Table~\ref{tab:purification-comparison-rvec} illustrates the AdvFAR, AdvFRR, and GenEER of the r-vector system integrated with the SSLR and various traditional filters. We observe that all the SSLR and filters can effectively purify adversarial samples, resulting in lower AdvFAR and AdvFRR than the ASV system without any defense methods. However, all of them negatively influence genuine samples, resulting in higher GenEER than the original ASV. 
A possible reason is that the SSLR models and filters cannot perfectly reconstruct the genuine samples, so some noise is inevitably added into the genuine samples, resulting in higher GenEER. 
However, as shown in Table~\ref{tab:purification-comparison-rvec}, our proposed SSLR models have better performance on both purifying the adversarial samples and maintaining the genuine samples than traditional filters. 
Similar observations on the x-vector system are illustrated in Table~\ref{tab:purification-comparison-xvec}.

\begin{table}[t]
\caption{The adversarial samples purification effectiveness comparison between the SSLR models and traditional filters. The ASV system is the r-vector system.}
\label{tab:purification-comparison-rvec}
\centering
\begin{tabular}{c|c|c|c}
\hline
\hline
        & AdvFAR (\%) & AdvFRR (\%) & GenEER (\%) \\
\hline
  NA        &       72.30 &     64.29   &    8.87  \\
\hline

  mean      &       29.74 &     29.00   &    27.59  \\

  median    &       30.67 &     29.65   &    27.30  \\

gaussian    &       32.16 &     32.68   &    30.80  \\

1*SSLR-C    &       22.86 &   25.97 &    22.59  \\
2*SSLR-C    &       29.74 &   32.90 &    29.50 \\

1*SSLR-CM  &       23.23 &    24.68 &    22.59 \\
2*SSLR-CM  &       26.58 &    29.00 &    27.30 \\

9*SSLR-TCM &      \textbf{20.82} &  \textbf{22.72}  &    \textbf{14.59}  \\
\hline
\hline
\end{tabular}
\end{table}

\begin{table}[th]
\caption{The adversarial samples purification effectiveness comparison between the SSLR models and traditional filters. The ASV system is the x-vector system.}
\label{tab:purification-comparison-xvec}
\centering
\begin{tabular}{c|c|c|c}
\hline
\hline
        & AdvFAR (\%) & AdvFRR (\%) & GenEER (\%) \\
\hline
  NA        &       67.29 &     82.25 &      6.59  \\
\hline

  mean      &       28.07 &     27.71 &     25.80  \\

  median   &       28.81 &    27.49 &     23.69  \\

gaussian    &       30.48 &   28.79 &     28.20  \\

1*SSLR-C    &       \textbf{7.43} &   61.90   &     19.50  \\
2*SSLR-C    &       14.31 &     54.76   &     26.69  \\

1*SSLR-CM  &       11.90 &    33.98   &     18.40  \\
2*SSLR-CM  &       21.75 &      34.42   &     25.19  \\

9*SSLR-TCM &      26.21 &   \textbf{27.06}  & \textbf{11.29}  \\

\hline
\hline
\end{tabular}
\end{table}

Also our experiments show that integrating ASV with either one SSLR-C or one SSLR-CM also decreases AdvFAR and AdvFRR, as shown in Table~\ref{tab:purification-comparison-rvec}. 
However, their negative influences on genuine samples are large, resulting in a large GenEER.
Integrating ASV with one or more SSLR-C or SSLR-CM will seriously degrade ASV systems' performance on genuine samples, so we won't use SSLR-C and SSLR-CM for defense.
We think that using channel masking during training will emphasize the imperfect reconstruction of SSLR models. The imperfect reconstruction is from the large distortions of the audio signal after passing through such models. But when combining channel masking with time masking during training, the SSLR models obtain the ability to do reconstruction by context information, thus alleviating the imperfect reconstruction. 
For SSLR-C and SSLR-CM, the negative effects induced by imperfect reconstruction outweighs the positive effect on alleviating adversarial noise when the cascading number of SSLR models increases.
Thus the performance of SSLR-C and SSLR-CM are not consistent with other SSLR models trained with time masking.
For SSLR-TCM, we consider cascading 9 blocks for performance comparison, as it achieves the balance on purifying adversarial samples and preserving the ASV performance on genuine samples. 
Similar observations on the x-vector system are illustrated in Table~\ref{tab:purification-comparison-xvec}.

\begin{figure}[t]
    \centering
    \includegraphics[width=0.45\textwidth]{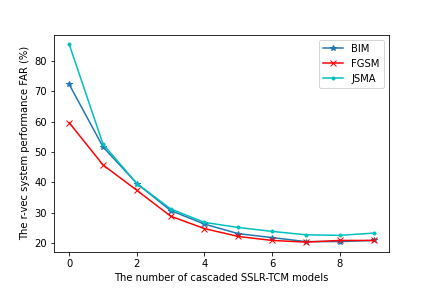}
    \caption{Purification performance under different attacks: the AdvFAR of the r-vector system under three attacks.}
    \label{fig:purification-all-attacks}
\end{figure}

\begin{figure}[t]
    \centering
    \includegraphics[width=0.45\textwidth]{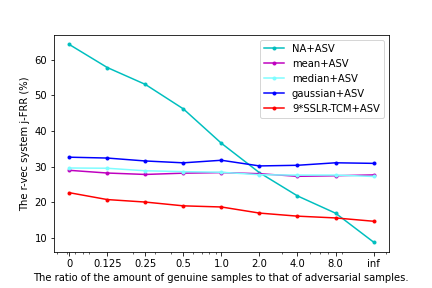}
    \caption{The j-FRR performance for the r-vector system integrated with different filters under a range of combination portion values.}
    \label{fig:purification-jfarjfrr}
\end{figure}

We also performed the proposed purification module on other attacks. 
As many kinds of SSLR model and evaluation metrics have been involved, this section only picks up the SSLR model with the best performance, i.e. SSLR-TCM. For evaluation metric, we will demonstrate the AdvFAR on the r-vector system under three attacks. Other results have similar trends.
The AdvFAR of the r-vector system under three attacks is illustrated in Fig.~\ref{fig:purification-all-attacks}. We observe that the AdvFAR dramatically decreases for all three attacks as the number of cascaded SSLR-TCM models increases, and the three decline curves have similar trends. This suggests that the proposed purification module can generalize to two other attacks, even though the defense method does not require any knowledge of the attack algorithms. This phenomenon shows an important advantage over previous attack-specific methods.

\begin{table*}[t]
\caption{The r-vector system performance under different conditions. (Each column refers to the system utilized to generate adversarial samples, while each row refers to the system for defense.)}
\label{tab:rvec-performance-under-diff-conditions}
\centering
\begin{tabular}{c|c|c|c|c|c|c}
\hline
\hline
\multirow{2}{*}{} & \multicolumn{2}{c|}{NA+ASV} & \multicolumn{2}{c|}{SSLR-TC+ASV} & \multicolumn{2}{c}{SSLR-TCM+ASV} \\
\cline{2-7}
                  & j-FAR (\%) & j-FRR (\%) & j-FAR (\%) & j-FRR (\%) & j-FAR (\%) & j-FRR (\%) \\
\hline
  NA+ASV        &       40.52    & 36.58      & 32.43    & 31.39 & 32.62       & 31.39      \\

  mean+ASV      &       28.72    & 28.36      & 28.81    & 28.36 & 28.72       & 28.57   \\

  median+ASV    &       28.90      & 28.46      & 28.90    & 28.25 & 29.37      & 28.46     \\

gaussian+ASV    &       31.51    & 31.82      & 31.88    & 31.71 & 31.69       & 31.60      \\

9*SSLR-TCM+ASV & \textbf{17.75}  & \textbf{18.72} & \textbf{20.35}  & \textbf{20.56} &  \textbf{23.14} &  \textbf{23.70}  \\

\hline
\hline
\end{tabular}
\end{table*}

Here we evaluate the j-FAR and j-FRR performance for both r-vector and x-vector system (totally 4 settings) under a range of combination portion values. These 4 settings have similar performance trends, and Fig.~\ref{fig:purification-jfarjfrr} illustrates the j-FRR for the r-vector system. In the condition of ASV-only (NA+ASV in Fig.~\ref{fig:purification-jfarjfrr}), we observe that j-FRR decreases as the portion of genuine samples increases. And all filters benefit ASV mostly in the area with high portion of adversarial samples. Specifically, our proposed SSLR models perform much better than traditional filters. However, as the portion of genuine samples increases, the benefits decrease. ASV-only begins to outperform ASV integrated with traditional filters at the ratio of around 2.0, and outperform ASV integrated with SSLR models after the ratio of 8.0. The reason is that as the portion of genuine samples increases, the negative influence of filters on genuine samples outweighs the positive influence on adversarial samples. Due to limited space, for the rest of this paper, we compute j-FAR and j-FRR by combining adversarial and genuine samples in equal proportion, unless otherwise specified.


\subsection{Attackers are aware of the SSLR models}
This section gives a case study to show the robustness of our adversarial perturbation purification module to a more severe attacking scenario where attackers have some knowledge about the SSLR. Specifically, we consider two scenarios, where 1) attackers are aware of the SSLR but do not know the specific masking strategies during training, 2) attackers know the specific masking strategies and can access the internal parameters of the SSLR. 
The SSLR model used for defense in this subsection is SSLR-TCM. To mimic such two scenarios, the SSLR models used for generating adversarial samples for Scenario 1) and 2) are SSLR-TC and SSLR-TCM, respectively. 
In this work, we only give a case study to cascade one SSLR model in front of ASV to generate adversarial samples for both Scenario 1 and 2. 

\begin{figure}[t]
    \centering
    \includegraphics[width=0.45\textwidth]{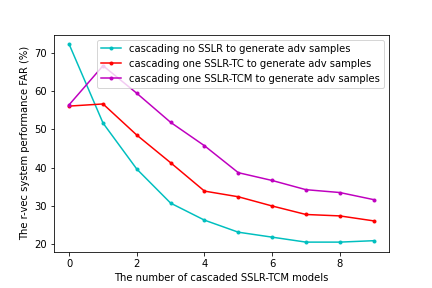}
    \caption{The AdvFAR performance trends for the r-vector system according to different knowledge of SSLR that attackers can access.}
    \label{fig:purification-some-knowledge-SSLR}
\end{figure}

This section compares the attack destructiveness in the above two scenarios and the normal scenario where attackers are unaware of the SSLR models. The AdvFAR and AdvFRR for the r-vector and x-vector systems (totally 4 settings) have similar trends, and Fig.~\ref{fig:purification-some-knowledge-SSLR} only demonstrates the AdvFAR of the r-vector system due to limited space.
By comparing the three lines in figures, we observe that as attackers can access more knowledge of the SSLR models, the generated adversarial samples become more effective. However, even if attackers can access the complete parameters of the defense SSLR (SSLR-TCM in our case), the attack destructiveness is still alleviated by integrating ASV with more SSLR models.

Considering a realistic scenario where adversarial and genuine samples are pooled together, we evaluate j-FAR and j-FRR of the ASV systems under three different scenarios. Table~\ref{tab:rvec-performance-under-diff-conditions} illustrates the performance of the r-vector system under three different scenarios. We also observe similar results on the x-vector system. 
From the table, we observe that our proposed SSLR outperforms traditional filters for both j-FAR and j-FRR in all three scenarios, which illustrates the effectiveness of SSLR on both purifying adversarial samples and also preserving ASV performance on genuine samples. Notice that in some scenarios, cascading a Gaussian filter may lead to worse performance on j-FRR than ASV only. The reason is that the Gaussian filter has larger negative influences on genuine samples, which outweighs the positive influences on adversarial samples.

\subsection{ASV fine-tuning}
\label{sec:asv-finetuning}
To alleviate the negative effects of the SSLR models on genuine samples, we make a further attempt to fine-tune the ASV models to better fit with the data generated by passing genuine samples through the SSLR models. 
Specifically, we pass the training data $D$, the development set of Voxceleb1, into the SSLR models (9 SSLR-TCM models) to generate $\widetilde{D}$, and use the combination of $D$ and $\widetilde{D}$ to fine-tune the ASV models several epochs. The fine-tune epochs are 3 epochs and 6 epochs for r-vector and x-vector, respectively. After fine-tuning, the negative effects are obviously reduced and the GenEERs are 11.99\% and 8.29\% for 9 SSLR-TCM incorporating with the r-vector and x-vector systems, respectively. In the meantime, it is also observed that the purification performance on adversarial samples is as effective as the previous one without fine-tuning.

\section{Experimental Results: Adversarial Perturbation Detection}
\label{sec:det-all}

\begin{figure}[t]
    \centering
    \includegraphics[width=0.45\textwidth]{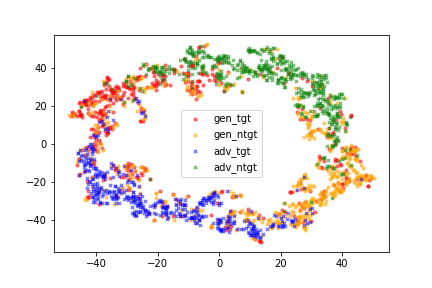}
    \caption{The t-SNE visualization of 10-dimensional scores for adversarial and genuine samples.}
    \label{fig:t-sne-visualization}
\end{figure}

In this section, we perform adversarial defense via a detection module. 
We found that ASV scores of adversarial samples change larger after integrating the ASV with the SSLR models than genuine samples. 
If we derive ASV scores by passing the data through a unique ASV integrated with different numbers of SSLR models and concatenating these scores as a vector, the distributions of the vectors for adversarial and genuine samples will be different. To demonstrate this property, given an utterance, we pass it through ASV integrated with 0 to 9 SSLR models to derive 10 scores as a 10-dimensional vector. 
Then, we use t-SNE to project this 10-dimensional vector into a 2-dimensional space and visualize both adversarial and genuine samples, as shown in Fig.~\ref{fig:t-sne-visualization}. 
This figure shows that adversarial and genuine samples can be well separated even in this 2-dimensional space. 
Besides, target and non-target trials are separated well. 
This phenomenon indicates that the vectors' statistics between adversarial and genuine samples are different, and motivates us to use the properties for detection.

As the scores of genuine and adversarial samples attain different statistical properties, we do experiments based on different orders of moments.
We demonstrate the detection EER ($EER_{det}$) for both r-vector and x-vector systems under three attack algorithms, as shown in Table~\ref{tab:det-eer-diff-moments}.
We observe that such a detection approach can detect all three attacks to some extent, even though the defense approach does not need any knowledge about the attack algorithms. This shows an advantage over previous approaches.
Specifically, we observe that odd orders of moments have a better detection performance over the even ones. The 3rd order moment and the 5th order moment achieve comparable detection results across all conditions.
For detection performance across different attack algorithms, we observe that attacks that achieve higher attack performance, i.e. higher EER in Table~\ref{tab:asv-performance}, are easier to be detected. For example, based on the r-vector system, JSMA achieves the highest attacking threat and FGSM shows the least attacking threat, accordingly, the $EER_{det}$ for JSMA is the lowest while that for FGSM is the highest. 
A possible explanation is that the more effective the attack is, the larger the variance of the scores passing through different systems consisting of different numbers of cascaded SSLR models will be. Hence, the more effective attack is kind of easier to be detected in our experiments.
Overall, Table~\ref{tab:det-eer-diff-moments} illustrates the detection accuracy for each condition is in a range, from 70.6\% to 86.4\%.


\begin{table}[t]
\caption{The detection equal error rate ($EER_{det} (\%)$) based on different statistical moments.}
\label{tab:det-eer-diff-moments}
\centering
\begin{tabular}{c|c|c|c|c|c|c}
\hline
\hline
\multirow{2}{*}{} & \multicolumn{3}{c|}{r-vec} & \multicolumn{3}{c}{x-vec} \\
\cline{2-7}
                  &  BIM & FGSM & JSMA         &  BIM & FGSM & JSMA \\
\hline
              2nd &  25.7          & 31.6          & 19.5          & 21.2          & 25.7          & 28.0 \\
              3rd &  \textbf{21.0} & 29.9          & \textbf{13.6} & \textbf{16.7} & 23.5          & \textbf{19.1} \\
              4th &  23.3          & 30.8          & 15.7          & 19.0          & 23.6          & 23.5 \\
              5th &  21.2          & \textbf{29.4} & 13.8          & 16.9          & \textbf{22.4} & 20.4 \\
\hline
\hline
\end{tabular}
\end{table}

\begin{table}[t]
\caption{The j-FAR and j-FRR of ASV and ASV integrated with the detection approach.}
\label{tab:det-tandem-performance-adv-all}
\centering
\begin{tabular}{c|c|c|c|c}
\hline
\hline
 \multirow{2}{*}{BIM}  &  \multicolumn{2}{c|}{r-vec} & \multicolumn{2}{c}{x-vec} \\
\cline{2-5}
                    & j-FAR (\%) & j-FRR (\%) & j-FAR (\%) & j-FRR (\%) \\
\hline
NA+ASV              & 39.08      & 8.87       & 31.21      & 6.71 \\
det+ASV             & 10.01      & 24.24      & 9.17       & 11.26 \\
\hline
\hline
 \multirow{2}{*}{FGSM}  &  \multicolumn{2}{c|}{r-vec} & \multicolumn{2}{c}{x-vec} \\
\cline{2-5}
                    & j-FAR (\%) & j-FRR (\%) & j-FAR (\%) & j-FRR (\%) \\
\hline
NA+ASV              & 37.19      & 8.87       & 30.36      & 6.71 \\
det+ASV             & 12.68      & 30.95      & 10.53      & 13.42 \\
\hline
\hline
 \multirow{2}{*}{JSMA}  &  \multicolumn{2}{c|}{r-vec} & \multicolumn{2}{c}{x-vec} \\
\cline{2-5}
                    & j-FAR (\%) & j-FRR (\%) & j-FAR (\%) & j-FRR (\%) \\
\hline
NA+ASV              & 39.60      & 8.87       & 31.27      &  6.71 \\
det+ASV             & 8.91       & 18.18      & 10.01    & 12.12 \\
\hline
\hline
\end{tabular}
\end{table}


We also conduct experiments based on the pool of adversarial and genuine samples and compare the j-FAR and j-FRR between ASV only and ASV with the detection module, as in (\ref{eq:det-jfar}) and (\ref{eq:det-jfrr}).
As shown in (\ref{eq:det-ntgt-set}) and (\ref{eq:det-tgt-set}), we treat all adversarial samples and genuine non-target samples as negative ones, while we treat genuine target samples as positive ones.
Since the 3rd and the 5th order moments achieve comparable results, as shown in Table~\ref{tab:det-eer-diff-moments}, we set the moment order as 5 for further experiments.
The j-FAR and j-FRR performance is demonstrated in Table~\ref{tab:det-tandem-performance-adv-all}. Across all three attacks, we observe that the j-FARs for both r-vector and x-vector systems dramatically decrease after integrating ASV with our detection approach, verifying the effectiveness of our detection module in such a tandem system. 
The j-FRRs for both r-vector and x-vector systems increase after applying the detection method. The reason is that our detection module will falsely reject some genuine target samples. 
However, we suggest that the positive effect on j-FARs outweighs the negative impact on j-FRRs. 
The increase of FAR is more dangerous than that of FRR in practical scenarios.
For example, in financial applications, the cost of a false alarm is usually higher than a false rejection.

\section{Conclusion}
In this work, we propose to harness self-supervised learning based reformer for adversarial defense from both adversarial perturbation purification and adversarial perturbation detection perspectives.
On the one hand, the adversarial perturbation purification effectively alleviates the adversarial noise in the inputs and recovers the clean counterpart from adversarial samples.
The SSLR degrades genuine GenEER very little after fine-tuning ASV (Sec.~\ref{sec:asv-finetuning}) – 3.12\% absolute and 1.70\% absolute for r-vector and x-vector systems, respectively.
This is a mild negative effect outweighed by the performance improvement of ASV systems under adversarial attack as shown in Figure~\ref{fig:purification-unaware-SSLR} - AdvFAR decreases from 72.30\% to 20.82\% and from 67.29\% to 26.21\% for r-vector and x-vector respectively, and AdvFRR decreases from 64.29\% to 22.73\% and from 82.25\% to 27.06\% for r-vector and x-vector respectively.
Also, our proposed method outperforms the traditional filters, which are firstly adopted for adversarial defense on ASV and set up as a baseline, not only on purifying the adversarial perturbation, but also on maintaining the performance of genuine samples.
On the other hand, the adversarial perturbation detection module helps detect the adversarial samples with an accuracy of 79.0\% and 83.3\% against BIM for r-vector and x-vector systems, respectively (Table~\ref{tab:det-eer-diff-moments}).
Our proposed method can also generalize to two other attack methods, as illustrated in  Figure~\ref{fig:purification-all-attacks} and Table~\ref{tab:det-eer-diff-moments}.

We also firstly benchmark the evaluation metrics for adversarial defense on ASV to make it easier for future works to conduct a fair comparison.
How to perfectly perform adversarial defense for ASV remains an open question, and we provide a potential direction for defense. 
We wish our method will provide inspiration for future researches about adversarial defense on ASV.

\section*{Acknowledgment}
This work is supported by the Ministry of Science and Technology of Taiwan (Project No. 110-2628-E-002-001), and HKSAR Government’s Research Grants Council General Research Fund (Project No. 14208718).


%

\begin{IEEEbiography}[{\includegraphics[width=1in,height=1.25in,clip,keepaspectratio]{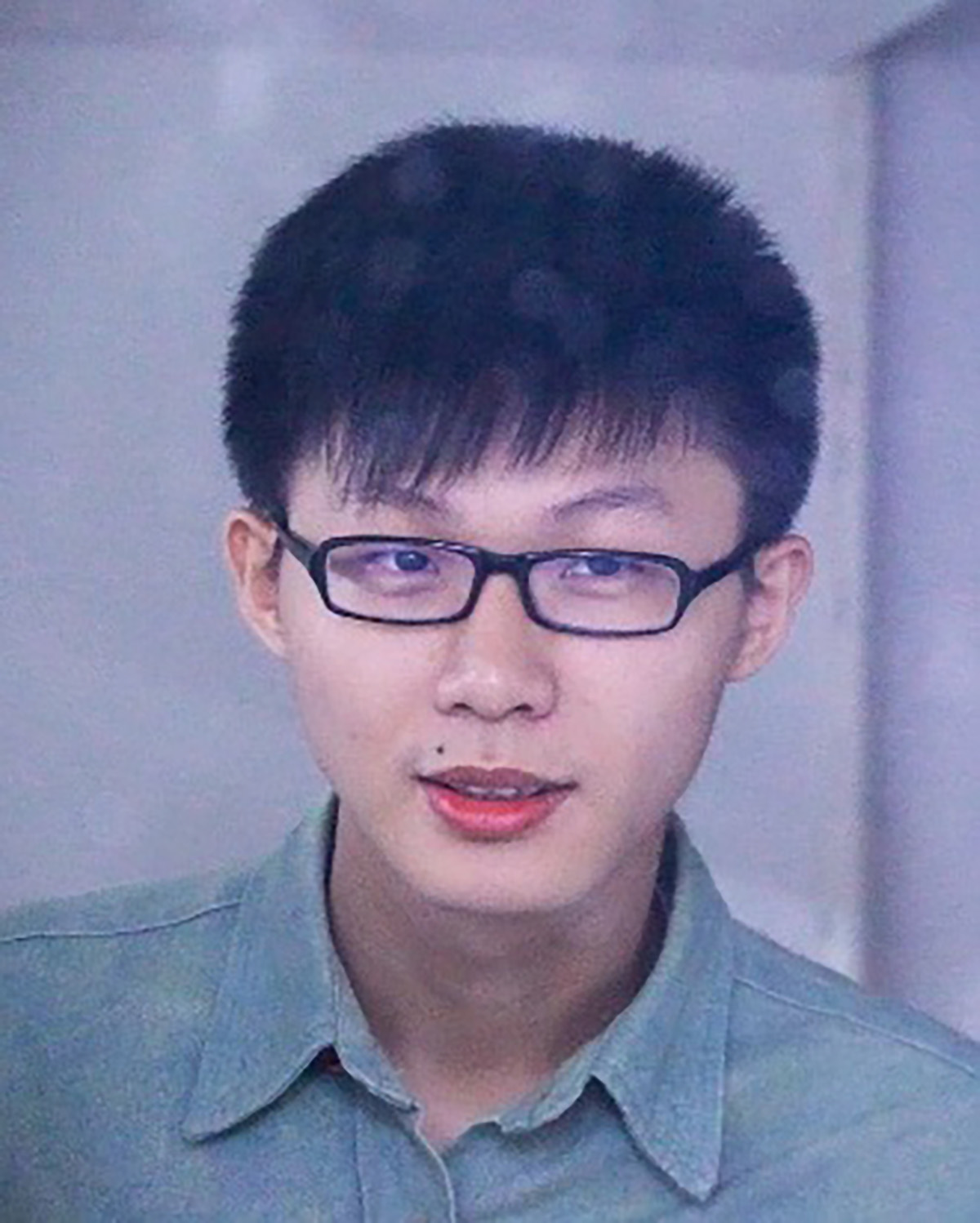}}]{Haibin Wu} Haibin Wu received the B.E. degree in Hunan University of Hunan, China. He is now a Ph.D. student in the Graduate Institute of Communication Engineering of National Taiwan University. His current research interests include automatic speaker verification, voice-print anti-spoofing, and self-supervised learning in the speech domain.
\end{IEEEbiography}

\begin{IEEEbiography}[{\includegraphics[width=1in,height=1.25in,clip,keepaspectratio]{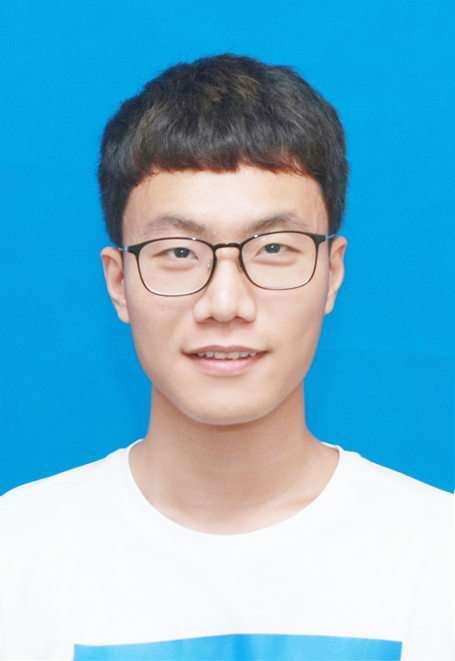}}]{Xu Li} Xu Li (Member, IEEE) received the B.E. degree in University of Science and Technology of China, Hefei, China, in 2017. Currently, he is a Ph.D. student at the Chinese University of Hong Kong. His current research interests include automatic speaker verification, voice-print anti-spoofing and language learning.
\end{IEEEbiography}

\begin{IEEEbiography}[{\includegraphics[width=1in,height=1.25in,clip,keepaspectratio]{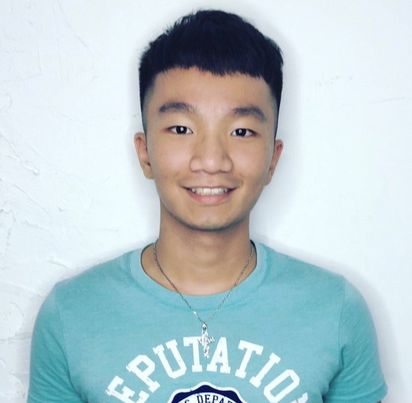}}]{Andy T. Liu}
Andy T. Liu received the bachelor’s degree in electrical engineering from National Taiwan University (NTU), Taipei, Taiwan, in 2018. He is currently working toward the Ph.D. degree with the Graduate Institute of Communication Engineering, NTU, supervised by Professor Hung-yi Lee. His research interests include self-supervised learning, few-shot learning, and machine learning in the speech and NLP domain.
\end{IEEEbiography}

\begin{IEEEbiography}[{\includegraphics[width=1in,height=1.25in,clip,keepaspectratio]{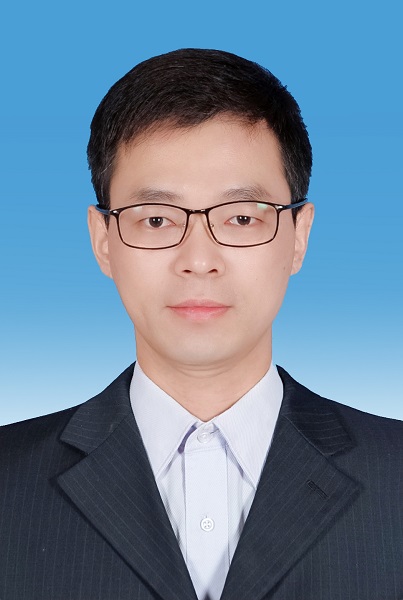}}]{Zhiyong Wu} Zhiyong Wu (Member, IEEE) received the B.S. and Ph.D. degrees in computer science and technology from Tsinghua University, Beijing, China, in 1999 and 2005, respectively. From 2005 to 2007, he was a Postdoctoral Fellow with the Department of Systems Engineering and EngineeringManagement, The Chinese University of Hong Kong (CUHK), Hong Kong. He then joined the Graduate School at Shenzhen (now Shenzhen International Graduate School), Tsinghua University, Shenzhen, China, and is currently an Associate Professor. He is also a Coordinator with Tsinghua-CUHK Joint Research Center for Media Sciences, Technologies and Systems. His research interests include intelligent speech interaction, more specially, speech processing, audiovisual bimodal modeling, text-to-audio-visual-speech synthesis, and natural language understanding and generation. He is a Member of International Speech Communication Association and China Computer Federation.
\end{IEEEbiography}

\begin{IEEEbiography}[{\includegraphics[width=1in,height=1.25in,clip,keepaspectratio]{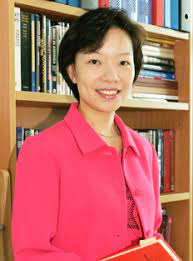}}]{Helen Meng} Helen Meng (Fellow, IEEE) received the B.S., M.S., and Ph.D. degrees in electrical engineering from the Massachusetts Institute of Technology, Cambridge, MA, USA. In 1998, she joined the Chinese University of Hong Kong, Hong Kong, where she is currently the Chair Professor with the Department of Systems Engineering and Engineering Management. She was the former Department Chairman and the Associate Dean of Research with the faculty of Engineering. Her research interests include human–computer interaction via multimodal and multilingual spoken language systems, spoken dialog systems, computer-aided pronunciation training, speech processing in assistive technologies, health-related applications, and big data decision analytics. She was the Editor-in-Chief of the IEEE TRANSACTIONS ON AUDIO, SPEECH AND LANGUAGE PROCESSING between 2009 and 2011. She was the recipient of the IEEE Signal Processing Society Leo L. Beranek Meritorious Service Award in 2019. She was also on the Elected Board Member of the International Speech Communication Association (ISCA) and an International Advisory Board Member. She is a ISCA, HKCS, and HKIE.
\end{IEEEbiography}

\begin{IEEEbiography}[{\includegraphics[width=1in,height=1.25in,clip,keepaspectratio]{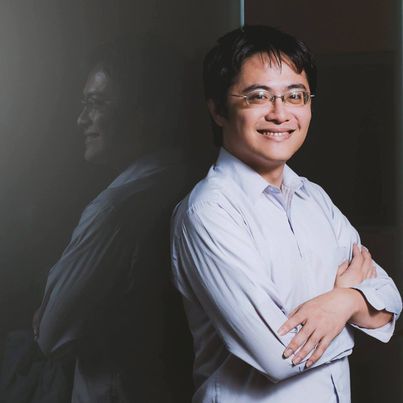}}]{Hung-yi Lee}
Hung-yi Lee received the M.S. and Ph.D. degrees from National Taiwan University, Taipei, Taiwan, in 2010 and 2012, respectively. From September 2012 to August 2013, he was a Postdoctoral Fellow with Research Center for Information Technology Innovation, Academia Sinica. From September 2013 to July 2014, he was a Visiting Scientist with the Spoken Language Systems Group, MIT Computer Science and Artificial Intelligence Laboratory. He is currently an Associate Professor with the Department of Electrical Engineering, National Taiwan University, Taipei, Taiwan with a joint appointment to the Department of Computer Science and Information Engineering. His research focuses on spoken language understanding, speech recognition, and machine learning.
\end{IEEEbiography}
\vfill








\end{document}